\documentclass[article,nojss,shortnames]{jss}

\usepackage[final]{changes}

\graphicspath{{Figures/}}
\usepackage{thumbpdf}
\usepackage{amsfonts,amstext,amsmath,amssymb,amsthm}
\usepackage{accents}
\usepackage{color}
\usepackage{rotating}
\usepackage{verbatim}
\usepackage{nicefrac}
\usepackage{adjustbox}
\usepackage{multirow}
\usepackage{numprint}

\usepackage{array}
\newcommand{\STAB}[1]{\begin{tabular}{@{}c@{}}#1\end{tabular}}

\newcommand{\THcite}[2]{\citeauthor{#2} (\citeyear{#2})}

\newcommand\norm[1]{\left\lVert#1\right\rVert}

\usepackage{booktabs}

\newcommand{\expit}{\text{logit}^{-1}}

\newcommand{\rY}{Y}
\newcommand{\rX}{\mX}
\newcommand{\rz}{z}
\newcommand{\ry}{y}
\newcommand{\rx}{\xvec}
\newcommand{\ru}{\uvec}

\newcommand{\samY}{\Xi}

\newcommand{\pZ}{F}

\newcommand{\pN}{\Phi}
\newcommand{\dN}{\phi}

\newcommand{\dZ}{f}

\newcommand{\h}{h}

\newcommand{\basisy}{\avec}

\newcommand{\parm}{\varthetavec}

\newcommand{\dimparm}{P}
\newcommand{\dimparmx}{Q}
\newcommand{\dimparmvar}{M}
\newcommand{\dimparmrand}{R}
\newcommand{\shiftparm}{\betavec}

\newcommand{\varparm}{\gammavec}
\newcommand{\eshiftparm}{\beta}

\newcommand{\ie}{\textit{i.e.}~}
\newcommand{\eg}{\textit{e.g.}~}

\newcommand{\RR}{\mathbb{R}}

\usepackage{dsfont}

 \DeclareMathOperator{\logit}{logit}

 \DeclareMathOperator{\trace}{tr}

 \DeclareMathOperator*{\argmax}{{arg\,max}}

 \DeclareMathOperator{\ND}{N}

\def \avec {\text{\boldmath$a$}}    \def \mA {\text{\boldmath$A$}}
    
    \def \mC {\text{\boldmath$C$}}
    \def \mD {\text{\boldmath$D$}}

    \def \mG {\text{\boldmath$G$}}
    
    \def \mI {\text{\boldmath$I$}}

    \def \mL {\text{\boldmath$L$}}

\def \pvec {\text{\boldmath$p$}}    
\def \qvec {\text{\boldmath$q$}}    
\def \rvec {\text{\boldmath$r$}}    \def \mR {\text{\boldmath$R$}}

\def \uvec {\text{\boldmath$u$}}    \def \mU {\text{\boldmath$U$}}
    \def \mV {\text{\boldmath$V$}}
\def \wvec {\text{\boldmath$w$}}    
\def \xvec {\text{\boldmath$x$}}    \def \mX {\text{\boldmath$X$}}
\def \yvec {\text{\boldmath$y$}}    \def \mY {\text{\boldmath$Y$}}
\def \zvec {\text{\boldmath$z$}}    \def \mZ {\text{\boldmath$Z$}}

\def \betavec         {\text{\boldmath$\beta$}}
\def \gammavec        {\text{\boldmath$\gamma$}}

\def \varepsilonvec   {\text{\boldmath$\varepsilon$}}

\def \varthetavec     {\text{\boldmath$\vartheta$}}

\def \muvec           {\text{\boldmath$\mu$}}

\def \phivec          {\text{\boldmath$\phi$}}

\def \mLambda  {\mathbf{\Lambda}}

\def \mSigma   {\mathbf{\Sigma}}

\def \mPhi     {\mathbf{\Phi}}

\def \nullvec {\mathbf{0}}
\def \onevec {\mathbf{1}}

\newcommand{\ubar}[1]{\underaccent{\bar}{#1}}

 \renewcommand{\Prob}{\mathbb{P}}

\author{Luisa Barbanti \\ Universit\"at Z\"urich \And Torsten Hothorn \\ Universit\"at Z\"urich}
\Plainauthor{Barbanti and Hothorn}

\title{A Transformation Perspective on Marginal and Conditional Models}
\Plaintitle{A Transformation Perspective on Marginal and Conditional Models}
\Shorttitle{A Transformation Perspective on Marginal and Conditional Models}

\Abstract{

Clustered observations are ubiquitous in controlled and observational
studies and arise naturally in multi-centre trials or longitudinal surveys. 
We present a novel model for the analysis of clustered observations where
the marginal distributions are described by a linear transformation model
and the correlations by a joint multivariate normal distribution.  The
joint model provides an analytic formula for the marginal distribution.
Owing to the richness of
transformation models, the techniques are applicable to any type of response
variable, including bounded, skewed, binary, ordinal, or survival responses. 
We \added{demonstrate how the common normal assumption for reaction times can be relaxed
in the sleep deprivation benchmark dataset and report marginal odds ratios
for the notoriously difficult toe nail data. We furthermore}
discuss the analysis of two clinical trials aiming at the estimation of
marginal treatment effects. In the first trial, pain was repeatedly assessed
on a bounded visual analog scale and marginal proportional-odds models are
presented. The second trial reported disease-free
survival in rectal cancer patients, where the marginal hazard ratio from Weibull
and Cox models is of special interest. \added{An empirical evaluation compares the
performance of the novel approach to general estimation equations for binary
responses and to conditional mixed-effects models for continuous responses.}
An implementation is available in the
\pkg{tram} add-on package to the \proglang{R} system and 
was benchmarked against established models in the literature.
 }

\Plainkeywords{conditional mixed-effects models, marginal models, marginal predictive
distributions, survival analysis, categorical data analysis}

\Address{
  Luisa Barbanti, Torsten Hothorn\\
  Institut f\"ur Epidemiologie, Biostatistik und Pr\"avention \\
  Universit\"at Z\"urich \\
  Hirschengraben 84, CH-8001 Z\"urich, Switzerland \\
  \texttt{Torsten.Hothorn@R-project.org}
}

\begin{document}

\section{Introduction}
In the context of the analysis of dependent data or clustered observations, 
many statistical \replaced{approaches}{models} for fitting conditional and marginal models have been 
studied.
Generalised mixed-models \citep[GLMMs,][]{Stroup_2012} \replaced{condition on
unobservable random effects and allow interpretation of covariate effects among
subjects sharing the same value of such a random effect. Conditional models
typically assume a specific random effects distribution and thus induce a
joint distribution from which marginal distributions can be derived either
analytically or by numerically integrating over random effects.}{
are typically fit via maximum-likelihood and allow for conditional inference
only.}
\replaced{Models
formulating marginal covariate associations without requiring a model for
the joint or conditional distribution}{On the other hand, marginal covariate associations} can be estimated by solving 
generalised estimating equations \citep[GEEs,][]{Zeger_Liang_1986,Zeger_Liang_1988}. 
Later, marginalised multilevel models \citep{Heagerty_1999,Heagerty_Zeger_2000}
were introduced providing a likelihood-based approach to estimate marginal 
coefficients in the framework of a conditional model. 
\THcite{Gory et al.}{Gory_Craigmile_MacEachern_2021} contribute a model definition allowing 
parameters estimated in a conditional model to be interpreted in a marginal fashion,
and \THcite{McGee and Stringer}{McGee_Stringer_2022} discuss marginal additive models for
potentially non-linear population-averaged associations, starting from
a generalised additive mixed-model framework.

Marginal predictive distributions with interpretable parameters are easy to
derive from normal linear mixed-effects models (LMMs) and binary probit GLMMs
as well as from some frailty models using a
copula representation \citep{Goethal_Janssen_2008}.  Regression coefficients
in other generalised linear mixed-effects or frailty models for non-normal
responses (binary logistic, Poisson, or Cox normal frailty models, for
example) only have a conditional interpretation, that is, given
unobservable \added{normal} random effects.  It is possible to obtain the marginal
\replaced{covariate effect}{distribution} by integrating out the \added{normal} random effects, however, the simple
interpretability of the fixed-effects regression coefficients is then lost. 
In contrast, marginal models allowing a marginal interpretation of effects
cannot be defined in an unambiguous way without the specification of a joint
distribution and we refer to \THcite{Lee and Nelder}{Lee_Nelder_2004} and
\THcite{Muff, Held, and Keller}{Muff_Held_Keller_2016} 
for a broader discussion of these issues.

Herein, we address the problem of formulating and estimating linear
transformation models \replaced{for the joint distribution}{in the presence} of cluster-correlated observations
arising, for example, in multi-centre trials or when a subject is repeatedly
examined over time.  In contrast to many methods in the mixed-effects and
frailty literature primarily aiming at explanation, that is, inference for
regression coefficients conditional on random effects in the presence of correlated
observations, the focus of this paper is on inference for marginal
distributions \added{which can be derived analytically from this novel joint
model.}  

Transformation models for correlated observations have been studied mostly
in the survival analysis context.  Parameter estimation is typically
performed by non-parametric maximum likelihood estimation
\citep{Cai_Wei_Wilcox_2000,Zeng_Gao_Lin_2017} where the transformation
function is only allowed to jump at distinct observed event times and is
conceptually understood as a infinite-dimensional nuisance parameter. 
\THcite{Lin, Lui, Xie, and Chen}{Lin_Luo_Xie_2017} even go a step further and 
propose a maximum rank
correlation estimator for estimating the regression coefficients
along with the mean of random effects while refraining from specifying
neither the transformation function nor a distribution for the random effects.  While this model
is extremely general, interpretation of the parameters is unclear and
predictive distributions cannot be derived.  Transformation models estimated
in a fully parametric way \citep{McLain_Ghosh_2013,Hothorn_Kneib_Buehlmann_2014,Hothorn_Moest_Buehlmann_2017,Klein_Hothorn_Kneib_2019}
are practically as flexible as semiparametrically estimated models yet
technically much easier to handle. \added{Therefore, various flavours of conditional
mixed-effects transformation models have been suggested
\citep{Manuguerra_Heller_2010,Garcia_Marder_Wang_2018,Tang_Wu_Chen_2018,Sun_Ding_2019,Tamasi_Crowther_Puhan_2022}.}
\deleted{We follow this approach by
choosing low-dimensional problem-specific parameterisations of the
transformation function.}

The results presented herein allow the formulation and estimation of models
for the joint multivariate distribution of clustered non-normally
distributed responses.  The marginal distributions obtained from the joint
distribution of observations in the same cluster feature directly interpretable
marginal effects.  Analytic formulae of the log-likelihood and the corresponding
score function for absolute continuous and potentially non-normal responses 
observed without censoring are available.  
For censored and discrete observations, evaluation
of the likelihood requires evaluation of multivariate normal probabilities,
however, in low dimensions such that the approximation error can be made
arbitrarily small in reasonable time. Applications from \deleted{two} different
domains presented in Section~\ref{sec:applications} highlight that the
methodology helps to unify models and inference procedures for clustered
observations across traditionally compartmentalised sub-disciplines of
statistics. \added{An empirical evaluation of marginal effects and
distributions estimated by this novel approach is presented in
Section~\ref{sec:sim}.} 

\section{Methods} \label{sec:methods}

Linear transformation models for the conditional distribution function
\begin{eqnarray} \label{fm:trafo}
\Prob(\rY \le \ry \mid \rX = \rx) = \pZ\left(\h(\ry) - \rx^\top \shiftparm\right)
\end{eqnarray}
of some univariate and at least ordered response $\rY \in \samY$ given a configuration
$\rx$ of covariates $\rX$ are defined by three objects: An
``inverse link function'' $\pZ$, a linear predictor $\rx^\top \shiftparm$
with regression coefficients $\shiftparm$ \emph{excluding} an intercept, and
a monotone non-decreasing transformation, or ``intercept'', function $\h$. 
Only $\shiftparm$ and $\h$ are unknowns to be estimated from data whereas
$\pZ$ defines the scale linearity of the effects is assumed upon.  This
model class covers many prominent regression models, such as normal,
log-normal, Weibull, or Cox models for absolute continuous responses, binary
models with different link functions, proportional odds and hazards
cumulative models for ordered responses, and many less well-known or even
novel models \citep{Hothorn_Moest_Buehlmann_2017}.
A summary of the possible inverse link functions $\pZ$ and of the corresponding
interpretation of the linear predictor $\rx^\top \shiftparm$ is given in 
Table~\ref{tab:links}.
This flexible modelling framework \replaced{covers}{allows the formulation
of} probabilistic index
models \citep{Thas_Neve_Clement_2012}, \ie models allowing \deleted{for an} interpretation of the effect size in terms
the probability that a randomly selected
subject has an outcome greater than the one of another randomly selected subject,
given that the covariate values for both subjects are known (for instance, whether
a subject received treatment or not). We illustrate the usefulness of this
quantity in Sections~\ref{sec:neck_app} and~\ref{sec:weibull_app}.

\begin{table}
\centering
\begin{tabular}{lll} \hline
\rule{0pt}{15pt}
$\pZ^{-1}(z)$ & $\pZ(z)$ & 
Interpretation of $\rx^\top \shiftparm$ \\ \hline \hline
probit & $\Phi_{0,1}(z)$ & conditional mean \\
& Standard normal & $\mathbb{E}(\h(\ry) \mid \rx) = \rx^\top \shiftparm$ \\
logit & $\expit(z) = \frac{1}{1 + \exp(-z)}$ & log-odds ratio \\
& Standard logistic & $\frac{\pZ(\h(\ry) \mid \rx)}{1 - \pZ(\h(\ry) \mid \rx)} = 
\exp(-\rx^\top \shiftparm) \frac{\pZ(\h(\ry))}{1 - \pZ(\h(\ry))}$ \\
cloglog & cloglog$^{-1}(z) = 1 - \exp(-\exp(z))$ &  log-hazard ratio\\
& Gompertz/Min. Extreme Value & $1 - \pZ(\h(\ry) \mid \rx) = 
\left(1 - \pZ(\h(\ry)) \right)^{\exp(-\rx^\top \shiftparm)}$\\
loglog & loglog$^{-1}(z) = \exp(-\exp(-z))$ & log-reverse time hazard ratio\\
&  Gumbel/Max. Extreme Value & $\pZ(\h(\ry) \mid \rx) = 
\pZ(\h(\ry))^{\exp(-\rx^\top \shiftparm)}$ \\ \hline
\end{tabular}
\caption{Interpretation of the linear predictor $\rx^\top \shiftparm$ under
different inverse link functions $\pZ$. In practice, many models are known
with respect to the link function $\pZ^{-1}$, which we report accordingly.
We denote the baseline ($\rx^\top \shiftparm = 0$) cumulative distribution 
function by $\pZ(\h(\ry))$
and the conditional cumulative distribution function by $\pZ(\h(\ry) \mid \rx)$.
\label{tab:links}}
\end{table}

For clustered or longitudinal data, we observe multiple values of the
response $\rY$ for each observational unit (clusters or subjects)
\added{whose interdependencies are not reflected in model~(\ref{fm:trafo}).
Adding, in analogy to GLMMs, a random effects term $\ru^\top \rvec$
to the linear predictor in~(\ref{fm:trafo}) defines mixed-effects transformation models}
\begin{align}
\Prob_{\rY}(Y \leq y \mid \rx, \ru, \rvec) = \pZ(\h(\ry) - 
\rx^\top \shiftparm - \ru^\top \rvec). \label{fm:tmME} \tag{tramME}
\end{align}
\added{When the random effects follow a specific bridge distribution to $\pZ$,
that is, the normal distribution for $\pZ = \Phi$, the stable distribution when $\pZ = \text{cloglog}^{-1}$
\citep{Aalen_2008}, or the distribution derived by \THcite{Wand and Louis}{Wang_Louis_2003} for $\pZ = \expit$,
marginal distributions can be derived. Neither the likelihood nor
marginal distributions, and thus a marginal interpretation of $\shiftparm$,
are available in closed form when the model is formulated differently,
especially when normal random effects are coupled with $\pZ \neq \Phi$
\citep{Tamasi_Crowther_Puhan_2022}.}

\subsection{Joint Transformation Models}

\replaced{To address these issues,}{In what follows,} we present a novel 
transformation model for the joint distribution that provides simple 
analytic expressions for marginal predictive distributions of the form 
(\ref{fm:trafo}). In this setup, $i = 1, \dots, N$ independent observational units, each consisting of
$N_i$ correlated observations of the response $\mY_i = (\rY_{i1}, \dots,
\rY_{iN_i})^\top \in \samY^{N_i}$, are available for estimating the joint
distribution. While refraining to specify a certain
parametric joint multivariate distribution for $\mY_i$, we assume that
probabilities on the scale of a suitable \emph{transformation} of
$\mY_i$ can be evaluated using a multivariate normal distribution whose
structured covariance matrix captures the correlations between the
transformed elements of $\mY_i$.  The aim of this paper is to simultaneously
estimate the transformation, regression coefficients, and the structured
covariance from data using models which emphasise predictive distributions
and parameter interpretability.

The non-decreasing transformation function $\h:\samY \rightarrow \RR$ is
applied element-wise to the response vector $\h_{N_i}(\mY_i) =
(\h(\rY_{i1}), \dots, \h(\rY_{iN_i}))^\top$ \added{ensuring that the same
transformation is applied to all $N_i$ observations.} Together with $\mY_i$, one
observes a corresponding \deleted{fixed effects design} matrix $\mX_i = (\rx_{i1} \mid
\dots \mid \rx_{iN_i})^\top \in \RR^{N_i \times \dimparmx}$ of full rank
\added{containing treatment assignment or covariates whose corresponding
\replaced{regression coefficients}{effects} $\shiftparm$ are of interest. In addition, the design of the
experiment is described by a} \deleted{design} matrix $\mU_i = (\ru_{i1} \mid \dots \mid
\ru_{iN_i})^\top \in \RR^{N_i \times \dimparmrand}$.  \added{We exclusively
study setups with simple cluster assignment encoded in this matrix ($\mU_i =
(1)_{N_i, 1}$) or longitudinal data ($\ru_{ij} = (1,
t_{ij})$ indicating that $\rY_{ij}$ for the $i$th subject was observed for at time
$t_{ij}$)}.
We propose to study
models for the joint distribution function of $\mY_i$ given $\rX_i$ 
and $\mU_i$ of the form
\begin{eqnarray} \label{fm:model}
\Prob(\mY_i \le \yvec \mid \mX_i, \mU_i) = \mPhi_{\nullvec_{N_i}, \mSigma_i(\varparm)}\left(
  \mD_i(\varparm) \Phi^{-1}_{N_i}(\pZ_{N_i}\{\mD_i(\varparm)^{-1}[\h_{N_i}(\yvec) - \mX_i \shiftparm]\})
  \right).
\end{eqnarray}
Here, $\mPhi_{\nullvec_{N_i}, \mSigma_i(\varparm)}\left( \cdot \right)$
is the distribution function of an $N_i$-dimensional
normal \added{random vector} with mean vector zero and structured covariance matrix
\begin{eqnarray*}
\mSigma_i(\varparm) := \mU_i \mLambda(\varparm) \mLambda(\varparm)^\top \mU_i^\top + \mI_{N_i}
\end{eqnarray*}
as defined by the random effects design matrix and \added{an unstructured} Cholesky factor
$\mLambda(\varparm) \in \RR^{\dimparmrand \times \dimparmrand}$ depending on
\added{unknown} variance parameters $\varparm \in \RR^{\nicefrac{\dimparmrand (\dimparmrand
+ 1)}{2}}$; $\mI_{N_i}$ denotes the $N_i \times N_i$ identity matrix.
We isolate the square roots of the diagonal elements of $\mSigma_i(\varparm)$ in the
matrix $\mD_i(\varparm) =
\text{diag}(\mSigma_i(\varparm))^{\nicefrac{1}{2}} \cdot \mI_{N_i} = \text{diag}(\mU_i
\mLambda(\varparm) \mLambda(\varparm)^\top \mU_i^\top +
\mI_{N_i})^{\nicefrac{1}{2}} \cdot \mI_{N_i}$.
A positive-semidefinite covariance matrix $\mSigma_i(\varparm)$
is given under the constraint $\text{diag}(\mLambda(\varparm)) \ge
\nullvec_R$. 
\added{For the simple model with $\mU_i = (1, \dots, 1)^\top$,
we have $\mLambda(\varparm) = \gamma_1$, $\mSigma_i(\varparm) =
(\gamma_1^2)_{N_i, N_i} + \mI_{N_i}$, and $\mD_i(\varparm)^{-1} =
(\gamma_1^2 + 1)^{-1} \cdot \mI_{N_i}$ is a scaling factor to the transformation function $\h$ 
and regression coefficients $\shiftparm$ which is instrumental for
the derivation of marginal distributions. In the longitudinal setup,
$\mLambda = \left( \begin{array}{cc}\gamma_1 & 0 \\
                                    \gamma_2 & \gamma_3 \end{array}
\right)$ 
and the covariance $\mSigma_i(\varparm)_{j,\jmath}$ depends on the
observation times $t_{ij}$ and $t_{i\jmath}$.} The
key component is the shifted transformation $\h_{N_i}(\yvec) - \mX_i
\shiftparm$ modelling the impact of the \replaced{regression coefficients}{fixed
effects} on the transformed
scale.

The transformation function $\h$, the regression coefficients $\shiftparm$,
and the variance parameters $\varparm$ are unknowns to be estimated from
data.  In (\ref{fm:model}), $\Phi_{N_i}^{-1}(\pvec) = (\Phi^{-1}(p_1), \dots,
\Phi^{-1}(p_{N_i}))^\top$ applies the quantile function $\Phi^{-1}$ of the
standard normal element-wise to some vector of probabilities $\pvec = (p_1,
\dots, p_{N_i})^\top \in (0,1)^{N_i}$.  Furthermore, $\pZ: \RR \rightarrow
\RR$ is an a priori defined cumulative distribution function of some
absolute continuous distribution with log-concave density $\dZ$; $\pZ_{N_i}$
and $\dZ_{N_i}$ are the element-wise applications of $\pZ$ and $\dZ$,
respectively.  

\begin{figure}[t]
\begin{center}
\includegraphics{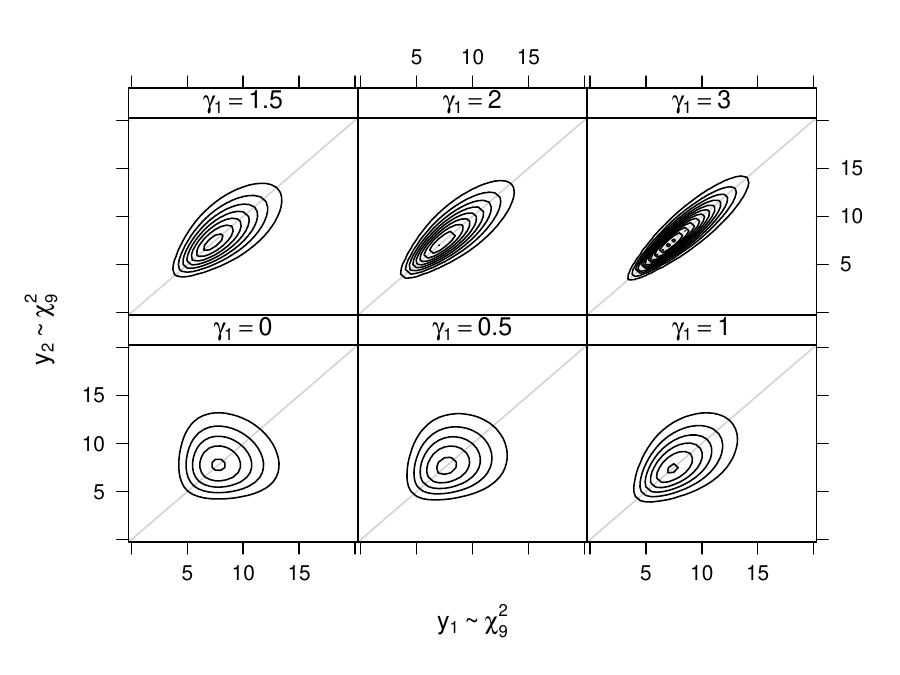}
\caption{Illustration. Bivariate joint density of an unconditional logistic ($\pZ = \logit^{-1}$) transformation model 
         for repeated measures (cluster size $N_i \equiv 2$, $\mU_i = (1, 1)^\top$ and 
         $\mSigma_i = \gamma_1^2 \mU_i \mU_i^\top + \mI_2$) with transformation functions $\h_1 = \h_2 =
         \sqrt{1 + \gamma_1^2} \cdot \logit \circ \chi^2_9$ such that both
         marginal distributions follow the $\chi^2_9$ law. 
         For $\gamma_1 = 0$, observations within a cluster are independent,
         and their correlation increases with increasing values of $\gamma_1$.
         \label{fig:chisq_ex}}
\end{center}
\end{figure}

For absolute continuous responses $\mY_i \in \RR^{N_i}$, model
(\ref{fm:model}) implies that the latent variable
\begin{eqnarray} \label{eq:Z_i}
\mZ_i := \mD_i(\varparm) \Phi^{-1}_{N_i}(\pZ_{N_i}\{\mD_i(\varparm)^{-1}[\h_{N_i}(\mY_i) - \mX_i \shiftparm]\}) \in \RR^{N_i}
\end{eqnarray}
defined as an element-wise transformation of the observations $\mY_i$
follows a multivariate normal distribution $\mZ_i \sim
\ND_{N_i}\left(\nullvec_{N_i}, \mSigma_i(\varparm)\right)$.    
\added{The model is distribution-free in the sense that for a baseline
configuration (with $\mX_i \shiftparm = \nullvec_{N_i}$), such a transformation
into multivariate normality exists for all marginal distributions
\citep{Klein_Hothorn_Kneib_2019}. The model does, however, impose a certain
correlation structure through the choice of $\mU_i$. An example for the joint
distributions induced by increasing correlations among bivariate repeated 
measurements with skewed marginal
distributions is given Figure~\ref{fig:chisq_ex}.}

The key aspect of an implementation of model (\ref{fm:model}) is the
parameterisation of the transformation function as $\h_{N_i}(\yvec) =
\mA(\yvec) \parm$ where $\mA(\yvec) = (\basisy(\ry_1) \mid \dots \mid
\basisy(\ry_{N_i}))^\top \in \RR^{N_i \times \dimparm}$ is the matrix of
evaluated basis functions $\basisy: \samY \rightarrow \RR^\dimparm$. 
Choices of basis functions $\basisy$ are problem-specific and
several options are discussed in Section~\ref{sec:applications} and, in more
detail, in \THcite{Hothorn, M\"ost, and B\"uhlmann}{Hothorn_Moest_Buehlmann_2017} and
\THcite{Hothorn}{Hothorn_2018_JSS}.

\subsection{Connection to Normal Linear Mixed-effects Models}

We first consider the special case $\pZ = \Phi$, where the 
transformation of $\mY_i$ simplifies to 
$\mZ_i = \h_{N_i}(\mY_i) - \mX_i \shiftparm =  \mA(\mY_i) \parm - \mX_i \shiftparm$. 
Model~(\ref{fm:model}) contains the LMM as a
special case. In its standard notation, the LMM reads
\begin{align} 
\mY_i = \alpha + \mX_i \tilde{\shiftparm} + \mU_i \mR_i  + \sigma \varepsilonvec_i
\label{fm:normal} \tag{LMM}
\end{align}
with random effects $\mR_i \sim \ND_{\dimparmrand}(\nullvec_\dimparmrand,
\mG(\varparm))$, residuals $\varepsilonvec_i \sim \ND_{N_i}(\nullvec_{N_i},
\mI_{N_i})$ under the assumption $\mR_i \perp \varepsilonvec_i$, intercept 
$\alpha \in \RR$ and residual standard deviation $\sigma \in \RR^+$. 
\added{The matrices $\mX_i$ and $\mU_i$ are typically referred to as ``fixed
effects'' and ``random effects'' design matrices in the literature.}
This model can be reformulated as a model for the joint multivariate distribution
\begin{eqnarray} \label{fm:normaldist}
\mZ_i = \frac{\mY_i - \alpha - \mX_i \tilde{\shiftparm}}{\sigma} = \mU_i \sigma^{-1} \mR_i  + \varepsilonvec_i
\sim \ND_{N_i}\left(\nullvec_{N_i}, \mU_i \mLambda(\varparm) \mLambda(\varparm)^\top \mU_i^\top +
\mI_{N_i}\right)
\end{eqnarray}
based on the relative covariance factorisation $\sigma^{-2}\mG(\varparm) =
\mLambda(\varparm)\mLambda(\varparm)^\top \in \RR^{\dimparmrand \times
\dimparmrand}$.  This is model (\ref{fm:model}) with $\pZ = \Phi$, linear
transformation $\h_{N_i}(\mY_i) = (\sigma^{-1}(\rY_{i1} - \alpha), \dots,
\sigma^{-1}(\rY_{iN_i} - \alpha))^\top = \mA(\mY_i) \parm$ with linear basis
functions $\basisy(\ry) = (\ry, -1)^\top$ and parameters 
$\parm = (\sigma^{-1}, \alpha \sigma^{-1})^\top$, and finally fixed effects 
$\shiftparm = \sigma^{-1} \tilde{\shiftparm}$.

Using this notation, the conditional distribution
function of some element $\ry \in \samY$ of $\mY$, conditional on $\rx$, $\ru$, and 
unobservable random effects $\mR = \rvec$, is 
\begin{eqnarray*} 
\Prob(\rY \le \ry \mid \rx, \ru, \rvec) = 
  \Phi\left(\basisy(\ry)^\top \parm - \rx^\top \shiftparm - \sigma^{-1} \ru^\top \rvec\right).
\end{eqnarray*}
The marginal distribution of some element $\ry \in \samY$ of $\mY$, which is
still conditional on $\rx$ and $\ru$ but integrates over the random effects
$\mR$, can be obtained from the joint multivariate normal
(\ref{fm:normaldist}) as
\begin{eqnarray*}
\Prob(\rY \le \ry \mid \rx, \ru) = 
\Phi\left(\frac{\basisy(\ry)^\top \parm - \rx^\top \shiftparm}{\sqrt{\ru^\top 
\mLambda(\varparm) \mLambda(\varparm)^\top \ru + 1}} \right).
\end{eqnarray*}
The shrunken marginal fixed effects $\shiftparm / \sqrt{\ru^\top
\mLambda(\varparm) \mLambda(\varparm)^\top \ru + 1}$ were also described by
\THcite{Wu and Wang}{Wu_Wang_2019} in a Bayesian implementation of this model. 
Understanding the LMM as special case of a
transformation model allows to relax the normality assumption for $\mY_i$ by
introducing non-linear transformation functions $\h(\ry) = \basisy(\ry)^\top
\parm$ defined by a non-linear basis $\basisy$
\citep{Hothorn_Moest_Buehlmann_2017}. Section~\ref{subsec:BoxCoxME} 
contains a comparison of the two models. 
\added{Probit GLMMs for binary responses $\rY \in \samY = \{0, 1\}$ 
can also be understood as a special case of a
transformation model with intercept $\h(0) = \alpha$ and $\h(1) = \infty$.
Several implementations of such GLMMs are compared empirically to an
implementation motivated from a transformation model perspective in
Section~\ref{subsec:bin}.}

\subsection{Distinction from Generalised Mixed-effects and Frailty Models}

Two important extensions of the LMM include
GLMMs and frailty models.  For binary
responses \deleted{$\rY \in \samY = \{0, 1\}$}, the logistic GLMM
has the conditional, given \added{normal} random effects $\rvec$, interpretation
\begin{eqnarray*}
\Prob(\rY = 0 \mid \rx, \ru, \rvec) = \expit\left(\alpha + \rx^\top \shiftparm + \ru^\top
\rvec\right).
\end{eqnarray*}
In survival analysis with $\rY \in \samY = \RR^+$, a Weibull normal frailty model 
leads to the conditional interpretation
\begin{eqnarray*}
\Prob(\rY \le \ry \mid \rx, \ru, \rvec) =
\text{cloglog}^{-1}\left(\alpha_1 + \alpha_2 \log(\ry) + \rx^\top \shiftparm  + \ru^\top \rvec\right).
\end{eqnarray*}
\added{A normal frailty Cox model}
\begin{eqnarray*}
\Prob(\rY \le \ry \mid \rx, \ru, \rvec) =
\text{cloglog}^{-1}\left(\h(\ry) + \rx^\top \shiftparm  + \ru^\top \rvec\right).
\end{eqnarray*}
\added{replaces the log-linear transformation function of the Weibull model 
with a smooth log-cumulative hazard function $\h(\ry)$. All three models are 
special cases of mixed-effects transformation models~(\ref{fm:tmME}).}

\added{Assuming normal random effects $\ru$,} neither model can be understood in terms of model (\ref{fm:model}) and two
main difficulties are associated with these types of models assuming
additivity of the fixed and random effects on the log-odds ratio or
log-hazard ratio scales.  First, unlike in the LMM~(\ref{fm:normal}), there is no analytic expression for the marginal
distribution and thus a marginal interpretation of the fixed effects
$\shiftparm$ is difficult.  Second, evaluation of the likelihood typically relies on a
Laplace approximation of the integral with respect to the random effects'
distribution and problems with this approximation have been reported,
for example by \THcite{Ogden}{Ogden_2015}.  The novel multivariate transformation
model for clustered observations based on~(\ref{fm:model}) addresses both of
these issues as shall be explained in the next subsections.

\subsection{Transformation Models with Marginal Interpretation}
\label{subsec:trafo}

Simple analytic expressions for the marginal distribution are available
(also for $\pZ \neq \Phi$), independent of the choice of the basis function
$\basisy$, noting that the variance of the $j$th element of $\mZ_i$ (Equation~\ref{eq:Z_i})
is $\ru_{ij}^\top \mLambda(\varparm) \mLambda(\varparm)^\top \ru_{ij} + 1$.


The Gaussian copula distribution of Equation~\ref{fm:model} directly implies the
marginal distribution function in form of a marginal transformation model
(mtram):
\begin{align}
\Prob(\rY \le \ry \mid \rx, \ru) &=
\Phi\left(\frac{\sqrt{\ru^\top \mLambda(\varparm) \mLambda(\varparm)^\top \ru +
1}\Phi^{-1}\left(\pZ\left(\frac{\basisy(\ry)^\top \parm - \rx^\top \shiftparm}
{\sqrt{\ru^\top \mLambda(\varparm) \mLambda(\varparm)^\top \ru +
1}}\right)\right)}{\sqrt{\ru^\top \mLambda(\varparm) \mLambda(\varparm)^\top \ru +
1}}\right) \notag \\
&= \pZ\left(\frac{\basisy(\ry)^\top \parm - \rx^\top \shiftparm}{\sqrt{\ru^\top \mLambda(\varparm) \mLambda(\varparm)^\top \ru +
1}}\right). \label{fm:tm2} \tag{mtram}
\end{align}
%
In this model, the fixed effects $\shiftparm$ divided by $\sqrt{\ru^\top
\mLambda(\varparm) \mLambda(\varparm)^\top \ru + 1}$ are directly
interpretable given $\mU = \uvec$, for example as log-odds ratios ($\pZ = \expit$) or
log-hazard ratios ($\pZ = \text{cloglog}^{-1}$).  Because $\mLambda(\varparm)
\mLambda(\varparm)^\top$ is positive semidefinite, there might be a reduction
in effect size when comparing the fixed effects $\shiftparm$ from 
formula~(\ref{fm:model}) to the marginal effects $\shiftparm / \sqrt{\ru^\top
\mLambda(\varparm) \mLambda(\varparm)^\top \ru + 1}$ from model~(\ref{fm:tm2}).  
For \replaced{repeated measurements}{random intercept-only models} with
$\ru = 1$ we get a constant reduction by $1 / \sqrt{\gamma_1^2 + 1}$.  In
longitudinal models \deleted{with correlated random intercepts and random
slopes}, the marginal effect at time $t$ is $\shiftparm / \sqrt{\gamma_1^2 +
\gamma_1\gamma_2 t + (\gamma_2^2 + \gamma_3^2) t^2 + 1}$ because $\ru = (1,
t)^\top$.  For positively correlated random intercepts and random slopes 
(\ie $\gamma_2 > 0$), the marginal effect always decreases over time.

\subsection{The Likelihood Function}

For parameters $\parm, \shiftparm$, and $\varparm$, the 
log-likelihood contribution $\ell_i(\parm, \shiftparm, \varparm)$ of the
$i$th subject or cluster is based on the transformation
\begin{eqnarray} \label{fm:z}
\zvec(\yvec \mid \parm, \shiftparm, \varparm) = \mD_i(\varparm) 
  \Phi^{-1}_{N_i}(\pZ_{N_i}\{\mD_i(\varparm)^{-1}[\mA(\yvec)\parm - \mX_i
\shiftparm]\})
\end{eqnarray}
of some $\yvec \in \samY^{N_i}$.  

For discrete or interval-censored
observations $(\ubar{\yvec}_i, \bar{\yvec}_i] \subset \RR^{N_i}$ the
log-likelihood contribution is
\begin{eqnarray} \label{fm:dll}
\ell_i(\parm, \shiftparm, \varparm) & = & 
    \log \Prob\left(\ubar{\yvec}_i \le \mY_i < \bar{\yvec}_i\right) = 
    \log \Prob\left(\zvec(\ubar{\yvec}_i \mid \parm, \shiftparm, \varparm) \le \mZ_i <
\zvec(\bar{\yvec}_i \mid \parm, \shiftparm, \varparm)\right) \notag \\
& = & \log\left\{
  \mPhi_{\nullvec_{N_i}, \mSigma_i(\varparm)}
  \left[\zvec(\ubar{\yvec}_i \mid \parm, \shiftparm, \varparm),
                   \zvec(\bar{\yvec}_i \mid \parm, \shiftparm, \varparm)
                   \right] \right\}
\end{eqnarray}
where
\begin{eqnarray*}
\mPhi_{\nullvec_{N_i}, \mSigma_i(\varparm)}(\ubar{\zvec}, \bar{\zvec}) = 
\int_{\ubar{\zvec}}^{\bar{\zvec}}
\phivec_{\nullvec_{N_i}, \mSigma_i(\varparm)}(\zvec) \, d\zvec
\end{eqnarray*}
is the integral over the $N_i$-dimensional multivariate normal density
$\phivec_{N_i}$ with mean zero and covariance $\mSigma_i$.  The structure of
$\mSigma_i(\varparm)$ can be exploited to dramatically reduce the
dimensionality of the integration problem.  Applying the procedure by
\THcite{Marsaglia}{Marsaglia_1963}, one can reduce this $N_i$-dimensional integral to an
$\dimparmrand$-dimensional integral over the unit cube (see Appendix~\ref{app:cens}).

For continuous observations $\yvec \in \RR^{N_i}$, it is common practice
\added{\citep[Section 5,][]{Lindsey_1999}} to
approximate this log-likelihood by a log-density evaluated at the observations 
$\yvec_i$:
%

\begin{eqnarray}  \label{fm:cll}
\ell_i(\parm, \shiftparm, \varparm) & \approx &
-\frac{1}{2}\log \left|\mSigma_i(\varparm)\right| + \notag \\
& & - \frac{1}{2} \zvec(\yvec_i \mid \parm, \shiftparm, \varparm)^\top 
(\mSigma_i(\varparm)^{-1} - \mD_i(\varparm)^{-2})
\zvec(\yvec_i \mid \parm, \shiftparm, \varparm) + \notag \\
& & \quad \log_{N_i}(\dZ_{N_i} \{ \mD_i(\varparm)^{-1} [\mA(\yvec_i) \parm - \mX_i \shiftparm] \} )^\top
\onevec_{N_i} + \\
& & \quad \log_{N_i}(
\mA^\prime(\yvec_i) \parm)^\top \onevec_{N_i} \notag 
\end{eqnarray}
where the Cholesky factorisation $\mL_i(\varparm) \mL_i(\varparm)^\top =
\mSigma_i(\varparm)$ is utilised.  It should be noted that the exact
log-likelihood function (\ref{fm:dll}) does not require the precision matrix
$\mSigma_i(\varparm)^{-1}$ to be computed.  
In the above approximation,
$\log_{N_i}$ is the element-wise natural logarithm and $\dZ_{N_i}$ the
element-wise density of $\pZ$.  $\mA^\prime(\yvec_i)$ denotes the matrix of
evaluated derivatives $\basisy^\prime$ of the basis function $\basisy$.  The
log-likelihood of Equation~\ref{fm:cll} is derived in Appendix~\ref{app:cont}.

Using either log-likelihood, we obtain simultaneous maximum-likelihood 
estimates for all model parameters from
\begin{eqnarray*}
(\hat{\parm}_N, \hat{\shiftparm}_N, \hat{\varparm}_N) = \argmax_{(\parm, \shiftparm,
\varparm) \in \RR^{\dimparm + \dimparmx + \dimparmvar}} \sum_{i = 1}^N
\ell_i(\parm, \shiftparm, \varparm).
\end{eqnarray*}
Some models require additional constraints on $\parm$ to be implemented
\citep{Hothorn_Moest_Buehlmann_2017}. Analytic score functions for
all model parameters $\parm, \shiftparm$, and $\varparm$ are available (see
Appendix~\ref{app:cont}).  Score functions for the discrete or censored
likelihood (\ref{fm:dll}) and the observed Fisher information matrices for
both likelihoods are obtained numerically. The full parameterisation of $\h$
allows application of standard results for likelihood asymptotics
\citep{vdVaart_1998} to independent observations \citep{Hothorn_Moest_Buehlmann_2017}.
Because the model~(\ref{fm:model}) is a special case of the multivariate
transformation model of \THcite{Klein et al.}{Klein_Hothorn_Kneib_2019} where the transformation
function $\h$ and the fixed effects $\shiftparm$ are constrained to be the same
for all ``coordinates'' of the random vector $\mY_i$ (that is, observations in the
same cluster). Therefore, model~(\ref{fm:model}) benefits from the same asymptotic results
reported by \THcite{Klein et al.}{Klein_Hothorn_Kneib_2019}.


\section{Applications} \label{sec:applications}


In this section, we discuss \replaced{four}{two} potential applications of
marginally interpretable transformation models.  Data, numerical details,
and code reproducing the results are available from the Online Appendix
\citep{vign:tram}.  \deleted{This document contains additional \added{analyses and}
applications from different domains.} We start with \replaced{two}{a} head-to-head
comparisons where model~(\ref{fm:tm2}) suggested here can be estimated by already
existing software implementations of \added{mixed-effects probit models} for the purpose of validating the
implementation of model~(\ref{fm:tm2}) in the add-on package \pkg{tram}
\citep{pkg:tram} to the \proglang{R} system for statistical computing.

\subsection{Non-normal Mixed-effects Models} \label{subsec:BoxCoxME}

\added{The average reaction times to a specific task over several days of sleep
deprivation are given for $i = 1, \dots, N = 18$ subjects 
\citep{Belenky_Wesensten_Thorne_2003}.  The data are often used to illustrate
LMMs with correlated random intercepts and slopes of the form
(\ref{fm:normal})}
\begin{eqnarray} \label{fm:sleep_lmer}
\Prob(\text{Reaction time} \le \ry \mid \text{day}, i) = 
  \Phi\left(\frac{\ry - \alpha - \beta \text{day} - \alpha_i - \beta_i \text{day}}{\sigma}\right),
    (\alpha_i, \beta_i) \sim \ND_2(\nullvec, \mG(\varparm)).
\end{eqnarray}
\added{This conditional normal model can be estimated by maximising the
corresponding normal log-likelihood and distinct implementations of classical
normal linear mixed
models \citep[\ref{fm:normal}, package \pkg{lme4},][]{pkg:lme4}, 
conditional mixed-effects transformation models \citep[\ref{fm:tmME}, package
\pkg{tramME},][]{Tamasi_Hothorn_2021}, and marginal transformation
models \citep[\ref{fm:tm2}, package \pkg{tram},][]{pkg:tram} provide identical results 
(in-sample log-likelihood $-875.97$).}

\added{Because the reaction times can hardly be expected to follow a symmetric
distribution, we consider the non-normal conditional and marginal 
transformation model}
\begin{eqnarray} \label{fm:sleep_tram}
\Prob(\text{Reaction} \le \ry \mid \text{day}, i) =
  \Phi\left(\h(\ry) - \beta \text{day} - \alpha_i - \beta_i \text{day}\right),
    (\alpha_i, \beta_i) \sim \ND_2(\nullvec, \mG(\varparm))
\end{eqnarray}
\added{where a monotonically increasing transformation function $\h(\ry)$ is
allowed to deviate from linearity.
Such probit-type mixed-effects models have been studied before, \eg by merging a
Box-Cox power transformation $\h$ with a grid-search over REML estimates
\citep{Gurka_Edwards_2006}, a conditional likelihood
\citep{Hutmacher_French_2011}, or a grid-search maximising the profile
likelihood \citep{Maruo_Yamaguchi_2017}.  Recently, \THcite{Tang, Wu, and
Chen}{Tang_Wu_Chen_2018} and \THcite{Wu and Wang}{Wu_Wang_2019} 
proposed a monotone spline parameterisation of $\h$ in a Bayesian context.}

\added{We parameterise $\h(\ry) = \basisy(\ry)^\top \parm$ in terms of a
monotonically increasing polynomial in Bernstein form of order six
\citep{Hothorn_Moest_Buehlmann_2017}.  The conditional transformation model
\citep{Tamasi_Crowther_Puhan_2022} can be estimated by maximising a Laplace
approximation to the log-likelihood
\citep{Tamasi_Hothorn_2021} simultaneously with respect to
all parameters $\parm$, $\eshiftparm$, and $\varparm$.  Direct optimisation 
of the log-likelihood~(\ref{fm:cll}) for the 
marginal transformation model~(\ref{fm:tm2}) leads to identical results (log-likelihood
$-859.55$), because the conditional and marginal models are identical for
$\pZ = \Phi$ and the Laplace approximation is
very accurate in this case.  For
$\pZ \neq \Phi$, conditional and marginal transformation models differ, and numerical integration with respect to the normal random
effects is required when marginal distributions shall be obtained from a
conditional model. In contrast, the marginal transformation model~(\ref{fm:tm2})
provides a closed-form expression for marginal distributions for all choices
of $\pZ$. With $\pZ = \expit$, the log-likelihood of the marginal
model increases slightly ($-860.6377$).}

\begin{figure}[t]
\begin{center}
\includegraphics{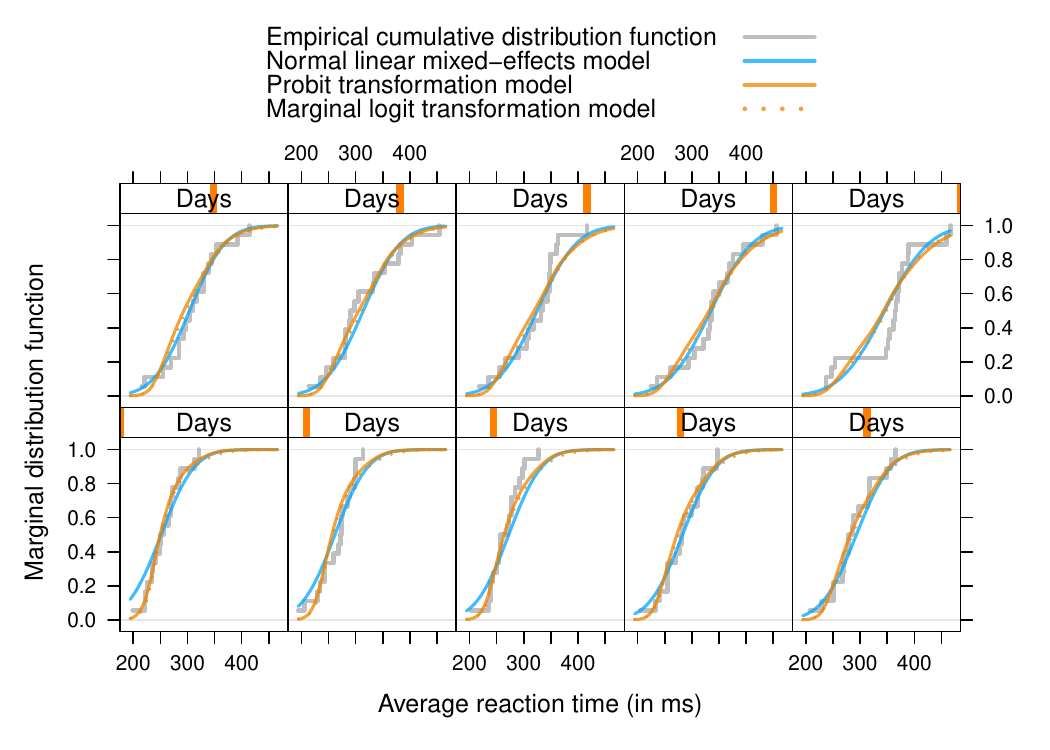}
\caption{Sleep deprivation. Marginal distribution of reaction times, separately for each day of
         study participation. The grey step-function corresponds to the
         empirical cumulative distribution function, the blue line to the
         marginal cumulative distribution of the normal linear mixed-effects model
         (\ref{fm:sleep_lmer}), estimated by the \code{lmer} function
         from package \pkg{lme4} \citep{pkg:lme4}, the solid yellowish line to the 
         probit transformation model (\ref{fm:sleep_tram}), and the dotted
         yellowish line to the logistic marginal transformation model. \label{fig:sleepstudy_ecdf}}
\end{center}
\end{figure}

\added{The daily marginal distribution
functions of normal and non-normal models are compared to the daily marginal empirical
cumulative distributions in Figure~\ref{fig:sleepstudy_ecdf}. Especially for
short reaction times early in the experiment, the non-normal transformation
models seem to fit the data better than the normal linear model. Between the
probit transformation model and the logistic marginal transformation model, only minor
discrepancies can be observed.}

\subsection{Binary Marginal Models} \label{subsec:bin}

\added{For a binary response $\ry \in \{0, 1\}$, the transformation $\h(\ry) =
\alpha$ reduces to a scalar intercept.  Thus, maximisation of the discrete
log-likelihood (\ref{fm:dll}) provides an alternative to commonly applied
approximations, such as Laplace or Adaptive Gauss-Hermite Quadrature, for
fitting conditional mixed-effects models. In addition, the possibility to interpret
parameters marginally also for $\pZ \neq \Phi$ asks for a comparison to
generalised estimation equations (GEEs).}

\added{We first compared different implementations of binary probit mixed-effects models
for the notoriously difficult to handle toe nail data
\citep{Backer_Vroey_1998} for which quasi-separation issues have been
reported \citep{Sauter_Held_2016}.
The ordinal response measuring toe nail infection was categorised to two
levels.  We were interested in binary probit models featuring fixed main and
interaction effects $\eshiftparm_1$, $\eshiftparm_2$, and $\eshiftparm_3$ of
treatment (itraconazole vs.~terbinafine) and time.  Subject-specific random
intercept models and models featuring correlated random intercepts and
slopes were estimated by the \code{glmer} function from package
\pkg{lme4} \citep{pkg:lme4}, by the \code{glmmTMB} function from package
\pkg{glmmTMB} \citep{pkg:glmmTMB}, and by direct maximisation of the exact
discrete log-likelihood (\ref{fm:dll}) given in Appendix~\ref{app:cens}.}

\added{The estimated model parameters, along with the discrete log-likelihood
(\ref{fm:dll}) evaluated at these parameters, are given in
Table~\ref{tab:toenail}.  For the random intercept models, AGQ, the
Laplace approximation in \pkg{glmmTMB}, and the
discrete log-likelihood gave the same results, the Laplace approximation
implemented in package \pkg{lme4}
seemed to fail.  It was not possible to apply the AGQ approach to
the random intercept / random slope model.  The two implementations of the
Laplace approximation in packages \pkg{lme4} and \pkg{glmmTMB} differed
for the random intercept but not for the random intercept / random slope
model.  The log-likelihood obtained by direct maximisation of
(\ref{fm:dll}) resulted in the best fitting model with the least extreme
parameter estimates.  Computing times for all procedures were comparable.}

\begin{table}
\begin{center}

\begin{tabular}{lrrrr|rrr} \\ \hline
& \multicolumn{4}{c|}{RI} & \multicolumn{3}{c}{RI + RS} \\
& \texttt{glmer} & \texttt{glmer} & \texttt{glmmTMB} &  & \texttt{glmer} & \texttt{glmmTMB} & \\
& L               & AGQ             & L & (\ref{fm:dll}) & L & L & (\ref{fm:dll}) \\ \hline \hline
$\alpha$ & -3.39 & -0.91 & -1.10 &  0.91 & -4.30 & -4.30 &  1.58
\\
$\eshiftparm_1$ & -0.03 & -0.11 & -0.17 & -0.11 &  0.05 &  0.05 &  0.27
\\
$\eshiftparm_2$ & -0.22 & -0.19 & -0.19 & -0.19 & -0.07 & -0.07 & -0.53
\\
$\eshiftparm_3$ & -0.07 & -0.06 & -0.06 & -0.06 & -0.23 & -0.23 & -0.18
\\
$\gamma_1$ &  4.57 &  2.12 &  2.10 &  2.11 & 10.88 & 11.01 &  5.22
\\
$\gamma_2$ &  0.00 &  0.00 &  0.00 &  0.00 & -1.64 & -1.68 & -0.37
\\
$\gamma_3$ &  0.00 &  0.00 &  0.00 &  0.00 &  0.79 &  0.83 &  0.53
\\
\hline
LogLik & -675.22&-637.34&-638.54&-637.34&-628.12&-630.65&-545.12 \\ 
Time (sec)   &  3.83& 2.40& 2.04& 2.20& 7.53& 3.44& 8.08 \\ \hline
\end{tabular}

\caption{Toe nail data. Binary probit models featuring fixed intercepts
$\alpha$, treatment effects $\eshiftparm_1$, time effects $\eshiftparm_2$,
and time-treatment interactions $\eshiftparm_3$ are compared.
Random intercept (RI) and
random intercept/random slope (RI + RS) models were estimated by the Laplace (L)
and Adaptive Gauss-Hermite Quadrature (AGQ) approximations to the likelihood (implemented in packages
\pkg{lme4} and \pkg{glmmTMB}). In addition, the exact discrete
log-likelihood (\ref{fm:dll}) was used for model fitting and evaluation (the
in-sample log-likelihood (\ref{fm:dll}) for all models and timings
of all procedures are given in the last two lines).
\label{tab:toenail}
}
\end{center}
\end{table}

\added{In a second step, a marginal transformation model with logit link was
compared to marginal odds ratios obtained from a GEE. We refitted published
GEE models for this data \citep[\proglang{SAS} results in Chapter 10,][]{Molenberghs_Verbeke_2005}
and noticed substantial differences indicating numerical instabilities for
this dataset (see Online Appendix). The monthly multiplicative treatment effect on the odds ratio 
scale was $0.91$ ($95\%$ confidence interval $0.83-1.00$) when a logistic
GEE with unstructured working correlation was estimated. The logistic
transformation model estimated the same parameter as $0.94$ ($95\%$ confidence interval
$0.89-0.99$). 
\citeauthor{Molenberghs_Verbeke_2005} (\citeyear{Molenberghs_Verbeke_2005},
p.~211) reported a GEE-based marginal odds
ratio of $0.89$ ($95\%$ confidence interval $0.81-0.98$, with
model-based standard errors and $\exp$-transformed Wald intervals). The
performance of GEEs and marginal transformation models are compared against
ground truth in a simulation experiment in Section~\ref{sec:sim}.}

\subsection{Models for Bounded Responses} \label{sec:neck_app}

\begin{figure}[t!]
\begin{center}
\includegraphics{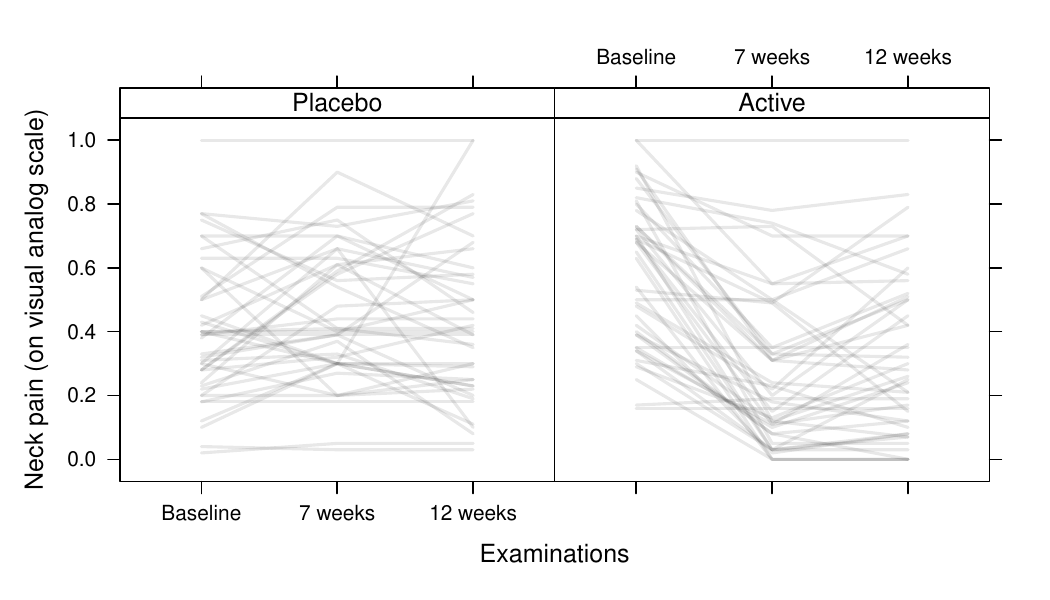}
\includegraphics{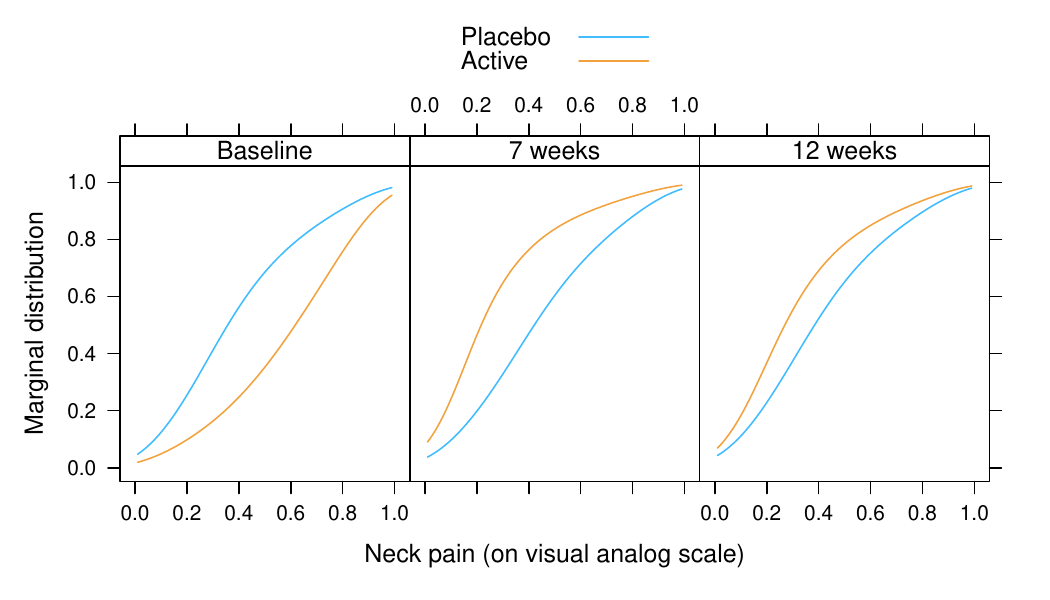}
\caption{Neck pain. Pain trajectories of $90$ subjects under active
         treatment or placebo evaluated at baseline, after $7$ and $12$ weeks 
         (top) and marginal distribution functions of neck pain at the 
         three different time points (bottom). These results were obtained
         from model (\ref{fm:tm2}) using $\pZ =
         \expit$ and a polynomial in Bernstein form $\h(\ry)$ on the unit
         interval. \label{fig:neck_pain}}
\end{center}
\end{figure}

\THcite{Chow and Heller}{Chow_Heller_2006} report on a randomised two-arm clinical trial
comparing a novel neck pain treatment to placebo.  Neck pain levels of $90$
subjects were assessed at baseline, after $7$, and after $12$ weeks
(complete trajectories are available for $84$ subjects) on a visual analog
scale.  \THcite{Manuguerra and Heller}{Manuguerra_Heller_2010} proposed a 
mixed-effects model for such a bounded response.  The fixed effects are interpretable as
log-odds ratios, conditional on random effects.  The data are presented in
the top panel of Figure~\ref{fig:neck_pain}.  A transformation model
(\ref{fm:tm2}) with $\pZ = \expit$ featuring a transformation function
$\h(\ry) = \basisy(\ry)^\top \parm$ defined by a polynomial in Bernstein
form of order six on the unit interval, and correlated random
intercept and random slope terms ($\ru = (1, t)$ for times $t = 0, 7, 12$ weeks)
is visualised by means of the corresponding marginal distribution functions
in the bottom panel of Figure~\ref{fig:neck_pain}.  Similar to the results
reported earlier \citep{Manuguerra_Heller_2010}, the model highlights more severe
pain in the active treatment group at baseline.  A positive treatment effect
can be inferred after $7$ weeks which seemed to level-off when subjects were
examined after $12$ weeks.  It is important to note that these results have
a marginal interpretation and that the model does not assume a specific
distribution of the response, such as a Beta distribution for example.

From the marginally interpretable transformation models, relevant quantities,
like the probabilistic index, can be derived \citep[Online Appendix,][]{vign:tram}.
In this application, the marginal probabilistic index is the probability that,
for a randomly selected patient in the treatment group, the neck pain score at 
time $t$ is higher than the score for a subject in the placebo group randomly 
selected at the same time point. 
We obtain a probability of $0.72$ ($95\%$ confidence interval $[0.58; 0.83]$) at baseline, $0.29$ ($95\%$ confidence interval
$[0.17; 0.43]$) after $7$ weeks and $0.38$ ($95\%$ confidence interval $[0.24; 0.54]$)
after $12$ weeks.

\subsection{Marginally Interpretable Survival Models} \label{sec:weibull_app}

The CAO/ARO/AIO-04 randomised clinical trial
\citep{Roedel_Graeven_Fietkau_2015} compared Oxaliplatin added to
fluorouracil-based preoperative chemoradiotherapy and postoperative
chemotherapy for rectal cancer patients to the same therapy using fluorouracil only.  
Patients were randomised in the two treatment arms by block
randomisation taking the study centre, the lymph node involvement (negative
vs.~positive), and tumour grading (T1-3 vs.~T4) into account.  The primary
endpoint was disease-free survival, defined as the time between
randomisation and non-radical surgery of the primary tumour (R2 resection),
locoregional recurrence after R0/1 resection, metastatic disease or
progression, or death from any cause, whichever occurred first.  The
observed responses are a mix of exact dates (time to death or incomplete
removal of the primary tumour), right-censoring (end of follow-up or
drop-out), and interval-censoring (local or distant metastases).
The conditional hazard ratio $0.79$ $(0.64,0.98)$ was reported as obtained from 
a Cox mixed-effects model with normal
random intercepts and without stratification fitted to right-censored
survival times \citep{Roedel_Graeven_Fietkau_2015}. This means that a rectal 
cancer patient treated with
the novel combination therapy benefits from a $21\%$ risk reduction
compared to a patient \emph{from the same block} treated with fluorouracil 
only.

We were interested in estimating a marginally interpretable treatment
\added{effect}
(acknowledging the fact that patients enrolled into the trial were
\emph{not} a random sample from all rectal cancer patients) based on a
marginally interpretable stratified (with respect to lymph node involvement
and tumour grading) Weibull model for clustered observations (blocks) in the
presence of interval-censored survival times.  This model can be formulated
by (\ref{fm:tm2}) choosing $\pZ = \text{cloglog}^{-1}$, $\basisy(\ry) = (1,
\log(\ry))^\top$, $\uvec = 1$ being the block indicator, and variance
parameter $\gamma_1$ (corresponding to the correlation structure of a random
intercept only model) as well as a treatment parameter $\eshiftparm$
(comparing the novum to fluorouracil only).  Stratification was implemented
by strata-specific parameters $\parm$ for each of the four strata.  It
should be noted that this model is not equivalent to a classical Weibull
normal frailty model.

A confidence interval for the marginal hazard ratio $\exp(\eshiftparm /
\sqrt{\gamma_1^2 + 1})$ was computed by simulating from the joint normal
distribution of $(\hat{\eshiftparm}, \hat{\gamma}_1)$.  With a relatively small 
$\hat{\gamma}_1 = 0.15$ (with standard error $0.13$), this resulted in a marginal
hazard ratio of $0.80$ ($95\%$ confidence interval $[0.65; 0.98]$), 
meaning that rectal cancer patients treated with
the combination therapy benefit from a $20\%$ risk reduction \emph{on average}.

By relaxing the Weibull assumption (log-linear transformation $\h$) to a Cox proportional hazards model
(nonlinear transformation $\h$), we obtain a
hazard ratio of $0.78$ ($95\%$ confidence interval $[0.64; 0.96]$)
and a marginal probabilistic index of
$0.56$ ($95\%$ confidence interval $[0.51; 0.61]$), meaning that over all study centres, a randomly selected patient
receiving Oxaliplatin has a $56\%$ probability of staying disease-free
longer than a randomly
selected patient receiving the standard treatment only, given that they both
have the same lymph node involvement and tumour grading.


\deleted{Linear transformation models~(\ref{fm:trafo}) have been extended to conditional
mixed-effects transformation models by including normal random effects
\mbox{\citep{Tamasi_Crowther_Puhan_2022}}
to allow the estimation of flexible regression models with correlated outcomes.
In this class of models, the conditional distribution
of the response given the fixed and random effects is given by <FORMULA>
%
%
where $\rvec \sim \ND(\nullvec, \mSigma)$.
Similarly to GLMMs, marginal effects cannot be read off directly from the output
of a mixed-effects transformation model. Moreover, the maximum-likelihood estimation
of the model parameters relies on the Laplace approximation of the
log-likelihood.}

\deleted{We compare the performance of conditional mixed-effects transformation
models~(\ref{fm:tmME}), in cases where the
model fitted is misspecified with respect to the data generating process.
We report results obtained when simulating a continuous response
in three
scenarios; details and code for the data generating process are outlined in the
Online Appendix.}

\deleted{First, we simulate from a transformation model with inverse link function
$\pZ = \pN$. In this setting, the conditional and the marginally interpretable
linear transformation model coincide, so we expect coefficients from
model~(\ref{fm:tmME}) and those from the model~(\ref{fm:tm2}) to coincide.
This is shown in Figure~\ref{fig:sim-probit_c}.
Then, we simulate from a conditional transformation model with inverse link
function $\pZ = \expit$. In this setting, model~(\ref{fm:tmME}) should outperform
model~(\ref{fm:tm2}).
Lastly, we simulate from a marginally interpretable transformation model with
inverse link function $\pZ = \expit$. In this setting, model~(\ref{fm:tm2})
should outperform the model~(\ref{fm:tmME}).
In Figure~\ref{fig:sim-tramME_c} we observe that the estimate of variance obtained
from model~(\ref{fm:tmME}) suffers from a slight bias.
This known and documented effect is due to the estimation of the parameters in a
maximum-likelihood framework. However, we clearly see that when the fitted model
does not reflect the data generating process on a theoretical level, the
marginal effects are still recovered correctly, but the variance is visibly
over- or underestimated. This is also the case for Figure~\ref{fig:sim-mtram_c}.
Similar effects are obtained in all three scenarios when simulating
interval-censored data.}

\section{Empirical Evaluation} \label{sec:sim}

\added{Practitioners interested in inference for marginal effects will likely apply
some form of GEE estimation when analysing a binary response, or might
integrate over random effects in a conditional mixed-effects model for more
complex response distributions. In this section, we assess the quality of
likelihood-based marginal transformation inference (model \ref{fm:tm2}) 
in comparison to GEEs for binary responses and to mixed-effects models for continuous
responses.}

\subsection{Data Generating Process}

\added{We simulate $N = 100$ clusters of five repeated measurements ($N_i = 5$ and
$\mU_i = (1, 1, 1, 1, 1)^\top$)
from a logistic model (\ref{fm:model}) with $\pZ = \logit^{-1}$ and 
transformation function $h = \sqrt{1 + \gamma_1^2} \cdot \text{logit} \circ
\chi^2_9$. The dependencies between repeated measurements in each cluster are described by
$\mSigma_i = (\gamma_1^2)_{5 \times 5} + \mI_5$. We are interested in
inference for the marginal effects $\muvec := (1 + \gamma_1^2)^{-\nicefrac{1}{2}}
\shiftparm$ for various values of $\gamma_1 \in \lbrace 0, 0.5, 1, 1.5, 2, 3 \rbrace$.
We simulated three uniform covariates $\rX$ and defined
$\shiftparm = (\beta_1, \beta_2, \beta_3)^\top = (0, 1, 2)^\top$. The
baseline distribution (with $\rx = (0, 0, 0)^\top$) induces the same marginal
$\chi^2_9$ laws for all five components with bivariate densities as depicted
in Figure~\ref{fig:chisq_ex}.}

\added{We report the mean-squared errors (MSEs) along with mean widths and coverages of $95\%$ confidence intervals 
for $\mu_p, p = 1, 2, 3$ based on \numprint{10000} simulation iterations in
Table~\ref{tab:sim}.}

\begin{table}[ht!]
\begin{center}
\begin{tabular}{l l c c c c c c c} \hline
\multicolumn{3}{c}{ } & $\gamma_1 = 0$ & $\gamma_1 = 0.5$ & $\gamma_1 = 1$ 
& $\gamma_1 = 1.5$ & $\gamma_1 = 2$ & $\gamma_1 = 3$ \\ \hline \hline
\multirow{9}{*}{\STAB{\rotatebox[origin=c]{90}{GEE (exchangeable)}}}
& \multirow{3}{*}{MSE}
 & $\mu_1$ & 0.109 & 0.104 & 0.087 & 0.070 & 0.061 & 0.050 \\
& & $\mu_2$ & 0.111 & 0.106 & 0.091 & 0.076 & 0.067 & 0.062 \\
& & $\mu_3$ & 0.120 & 0.114 & 0.101 & 0.093 & 0.091 & 0.094 \\ \cline{3-9}
& \multirow{3}{*}{CI width} 
 & $\mu_1$ & 1.273 & 1.247 & 1.141 & 1.035 & 0.958 & 0.868 \\
& & $\mu_2$ & 1.284 & 1.259 & 1.161 & 1.072 & 1.011 & 0.950 \\
& & $\mu_3$ & 1.317 & 1.295 & 1.219 & 1.169 & 1.153 & 1.165 \\ \cline{3-9}
& \multirow{3}{*}{Coverage} 
 & $\mu_1$ & 0.948 & 0.947 & 0.947 & 0.950 & 0.948 & 0.947 \\
& & $\mu_2$ & 0.944 & 0.946 & 0.945 & 0.947 & 0.950 & 0.945 \\
& & $\mu_3$ & 0.942 & 0.944 & 0.947 & 0.945 & 0.942 & 0.941 \\ \hline

\multirow{9}{*}{\STAB{\rotatebox[origin=c]{90}{mtram (binary)}}}
& \multirow{3}{*}{MSE} 
 & $\mu_1$ & 0.109 & 0.104 & 0.086 & 0.068 & 0.057 & 0.044 \\
& & $\mu_2$ & 0.110 & 0.106 & 0.091 & 0.074 & 0.064 & 0.055 \\
& & $\mu_3$ & 0.119 & 0.114 & 0.100 & 0.091 & 0.087 & 0.088 \\ \cline{3-9}
&\multirow{3}{*}{CI width} 
 & $\mu_1$ & 1.251 & 1.254 & 1.150 & 1.042 & 0.958 & 0.847 \\
& & $\mu_2$ & 1.276 & 1.268 & 1.172 & 1.079 & 1.014 & 0.942 \\
& & $\mu_3$ & 1.343 & 1.303 & 1.230 & 1.178 & 1.162 & 1.184 \\ \cline{3-9}
&\multirow{3}{*}{Coverage} 
 & $\mu_1$ & 0.953 & 0.951 & 0.949 & 0.953 & 0.953 & 0.952 \\
& & $\mu_2$ & 0.948 & 0.950 & 0.949 & 0.951 & 0.954 & 0.953 \\
& & $\mu_3$ & 0.947 & 0.948 & 0.950 & 0.951 & 0.953 & 0.955 \\ \hline

\multirow{9}{*}{\STAB{\rotatebox[origin=c]{90}{mtram (continuous)}}}
& \multirow{3}{*}{MSE} 
 & $\mu_1$ & 0.074 & 0.067 & 0.045 & 0.029 & 0.019 & 0.009 \\
& & $\mu_2$ & 0.079 & 0.070 & 0.048 & 0.033 & 0.024 & 0.015 \\
& & $\mu_3$ & 0.082 & 0.075 & 0.056 & 0.046 & 0.038 & 0.033 \\ \cline{3-9}
&\multirow{3}{*}{CI width} 
 & $\mu_1$ & 1.040 & 1.005 & 0.827 & 0.659 & 0.535 & 0.382 \\
& & $\mu_2$ & 1.061 & 1.020 & 0.853 & 0.705 & 0.602 & 0.485 \\
& & $\mu_3$ & 1.119 & 1.059 & 0.926 & 0.826 & 0.766 & 0.710 \\ \cline{3-9}
&\multirow{3}{*}{Coverage} 
 & $\mu_1$ & 0.949 & 0.949 & 0.948 & 0.945 & 0.947 & 0.948 \\
& & $\mu_2$ & 0.945 & 0.948 & 0.949 & 0.948 & 0.947 & 0.951 \\
& & $\mu_3$ & 0.949 & 0.947 & 0.947 & 0.945 & 0.951 & 0.950 \\  \hline
\end{tabular}
\caption{Simulations. MSE, widths and coverages of $95\%$ confidence
intervals for three marginal effects. For dichotomised binary responses,
results obtained from GEEs can be directly compared to results from marginal transformation
models (first two blocks). The last block reports results of marginal
transformation models fitted to continuous responses.}
\label{tab:sim}
\end{center}
\end{table}

\subsection{Binary Responses}

\added{Binary responses were generated by dichotomisation of the continuous
response at the overall median.  We fitted logistic GEEs with exchangeable
working correlation structure and computed estimates and confidence
intervals for all three marginal parameters $\mu_p, p = 1, 2, 3$. Results
are shown in the first block of Table~\ref{tab:sim}. In addition, marginal
transformation models were fitted to these binary responses. Joint
maximum-likelihood estimates of $\gamma_1$ and $\shiftparm$ were computed
from which we derived estimates and confidence intervals for the marginal
effects $\mu_p, p = 1, 2, 3$. We drew \numprint{10000} samples from the
asymptotic joint normal distribution of $\gamma_1$ and $\shiftparm$ to
derive confidence intervals for $\mu_p, p = 1, 2, 3$ in each simulation
iteration. These results in the
second block of Table~\ref{tab:sim} are practically equivalent to the
results reported for GEEs. For $\mu_2$ and $\mu_3$, the coverage of confidence
intervals computed from model~(\ref{fm:model}) were slightly closer to the
nominal $95\%$ level.}

\subsection{Continuous Responses}

\added{Marginal transformation models fitted to data on the original scale, \ie
without dichotomisation of the response, performed better in terms of
smaller MSEs and confidence interval widths (third block in
Table~\ref{tab:sim}). The coverage remained close to the nominal level.}

\added{In addition, we compared marginal transformation models for continuous
responses to two mixed-effects models: A normal linear mixed-effects model
(\ref{fm:normal}) and a conditional logistic mixed-effects transformation model
\citep[\ref{fm:tmME},][]{Tamasi_Hothorn_2021}. Unlike GEEs, these two additional competitors
are misspecified and one has to integrate over normal random effects to
obtain a marginal distribution given a specific configuration of $\rx$. For
the normal linear mixed-effects model, the marginal distribution is again
normal. Numerical integration was used to obtain marginal distributions from
the tramME model.}

\begin{figure}[t!]
\begin{center}
\includegraphics{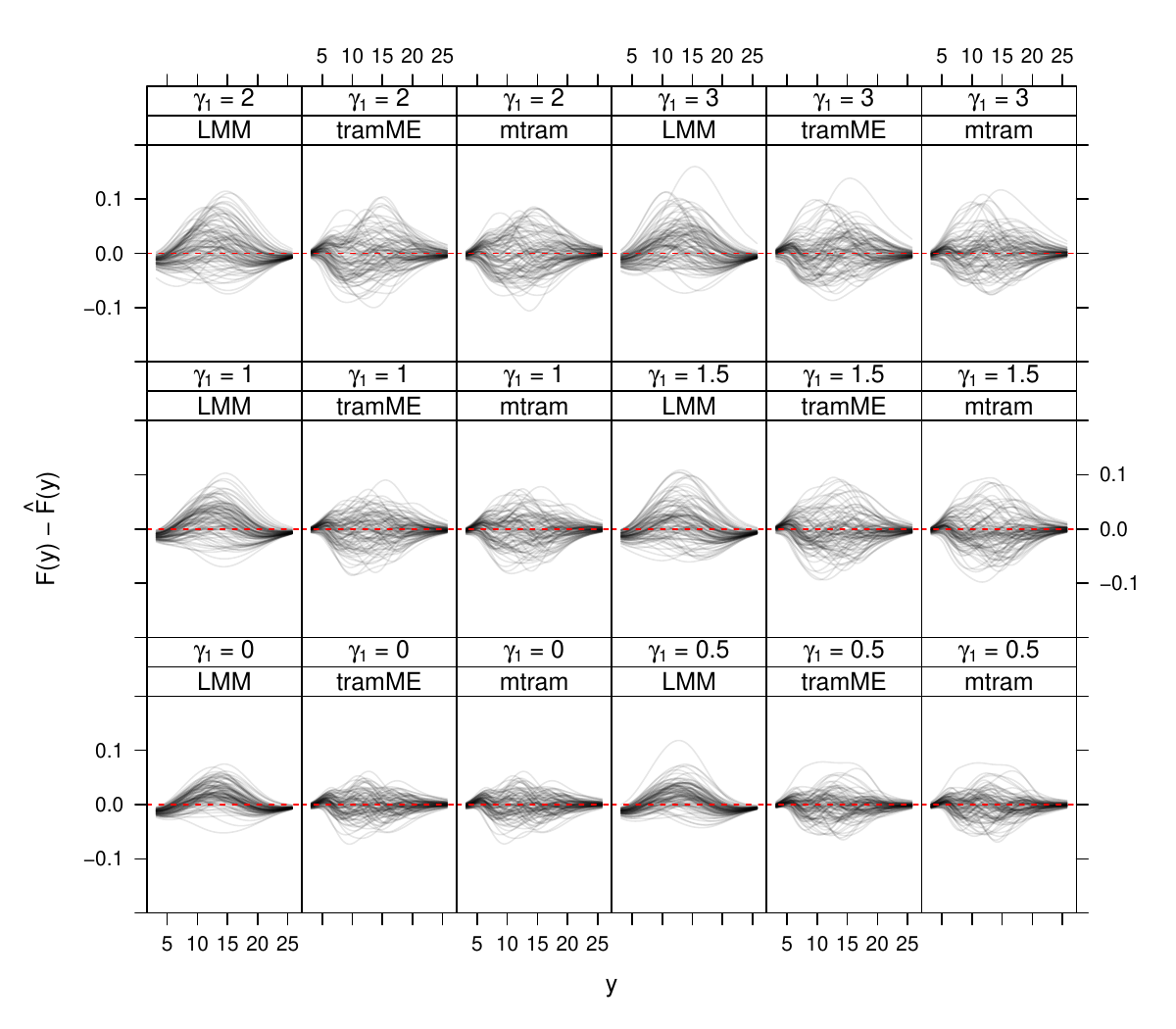}
\caption{Difference between the true and estimated marginal distribution
functions for a normal linear mixed-effects model (\ref{fm:normal}), a conditional
logistic mixed-effects transformation model (\ref{fm:tmME}) and a marginal
transformation model (\ref{fm:tm2}). For mixed-effects models,
the marginal distribution function was computed by integrating out the random
effects (analytically for LMM and numerically for tramME).
\label{fig:sim-cdfdiff}}
\end{center}
\end{figure}

\added{For a marginal transformation model~(\ref{fm:model}), a conditional logistic
mixed-effects transformation model with the same model complexity in terms
of parameters for the transformation function and for the shift parameters,
and a normal linear mixed-effects model, we derived the marginal
distribution conditional on $\rx = (.5, .5, .5)^\top$ for $100$ simulation
iterations and present the difference $F(\ry \mid \rx) - \hat{F}(\ry \mid
\rx)$ of the true and estimated marginal distribution functions for all
three procedures in Figure~\ref{fig:sim-cdfdiff}.  The normal linear
mixed-effects model (\ref{fm:normal}) lead to biased marginal distributions, simply
because the model is not able to adapt to the skewness of the marginal
distributions.  The results for the marginal (\ref{fm:tm2}) and conditional
(\ref{fm:tmME}) transformation models were surprisingly similar, especially for
smaller values of $\gamma_1$.  For $\gamma_1 = 0$ and thus independence
measurements, results are expected to be identical.  For $\gamma_1 = 3$, and
thus very large correlations among the five repeated measurements, the
estimated marginal distribution functions obtained from tramME seemed to be
slightly more biased than the marginal distribution functions obtained from
the marginal transformation model.}

\begin{figure}[t!]
\begin{center}
\includegraphics{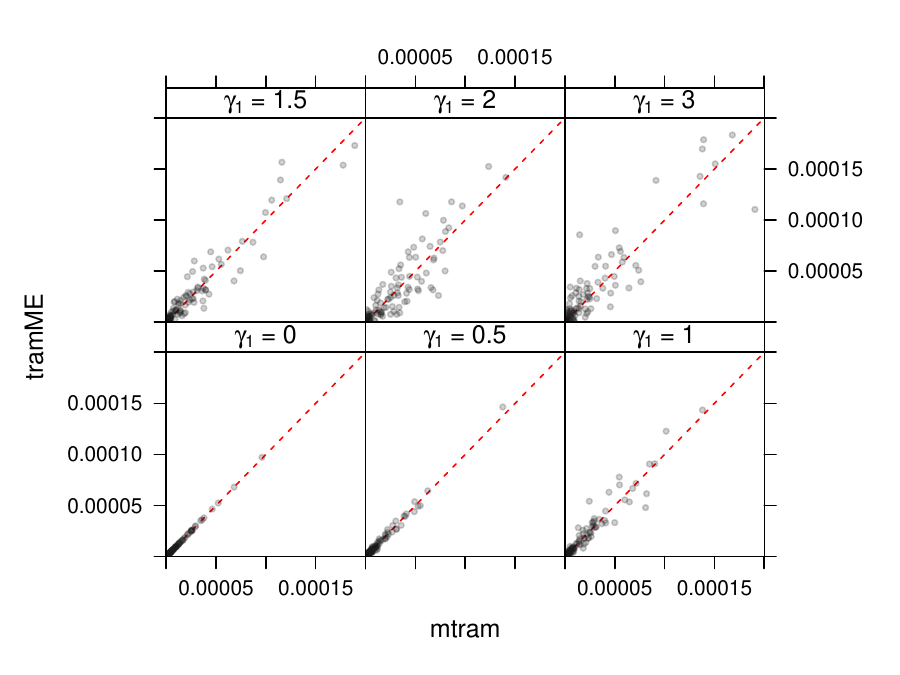}
\caption{Integrated mean-squared error between the true marginal distribution function
and the estimated marginal distribution function for a conditional logistic
mixed-effects transformation model (\ref{fm:tmME}) and a marginal transformation
model (\ref{fm:tm2}).
\label{fig:sim-imse}}
\end{center}
\end{figure}

\added{This impression is also supported in Figure~\ref{fig:sim-imse}, where the
integrated mean-squared error of the difference in distributions
$\int_{-\infty}^\infty \left(F(\ry \mid \rx) - \hat{F}(\ry \mid \rx) \right)^2 \cdot
f(\ry \mid \rx) \mathrm{d}y$ is presented for the conditional logistic mixed-effects
transformation model~(\ref{fm:tmME}) and the marginal transformation model~(\ref{fm:model}). 
For $\gamma_1 < 2$, the two
procedures performed very similar, for larger correlations the misspecified
tramME model exhibited slightly larger descrepancies between true and
estimated marginal distribution function. Of course, it is not possible to
derive marginal effects and corresponding confidence intervals from such 
numerically obtained marginal distributions.}

\section{Discussion}

There is a difference between a marginal and a marginally interpretable
model.  A marginal model, for example defined by generalised estimation
equations \citep{Zeger_Liang_1988}, does not specify the joint distribution. 
A marginally interpretable model is a model for the joint or conditional
(given random effects) distribution from which one can infer the marginal
distribution \citep{Lee_Nelder_2004}.  The models proposed here follow the
latter approach with the important distinctive feature that very simple
expressions for the marginal distribution function are available.  Thus,
there is no need to apply numerical integration to the joint or conditional
model formulation.  In our view, model (\ref{fm:tm2}) is especially
attractive because it allows the interpretation of scaled regression
coefficients as marginal effects acting on the marginal predictive
distribution in terms of a log-odds ratio or a log-hazard ratio, for example.
The Gaussian copula approach for obtaining marginally interpretable models has 
gained some interest in the last years 
\citep{Zhang_Charalambous_Foster_2021,Masarotto_Varin_2012}, however, the simple framework
of transformation models allows estimation for a wide range of responses without
encountering computational burdens or challenges that other methods typically do.
\deleted{The Online Appendix illustrates this point discussing
additional applications from different domains.}

Naturally, the questions arises which model is to be preferred: a marginal, a
conditional, or a marginally interpretable one? In this case, the ``right'' model
is not the model which most closely reflects the data generating process, which
is usually unknown, but rather the model that allows the user to answer the 
research question at hand by interpreting the estimated parameters, as 
\THcite{McGee and Stringer}{McGee_Stringer_2022} point out. An advantage of transformation models is
that besides allowing for interpretation of the fixed-effects on a marginal level,
they also yield valid models for the whole marginal
distribution~(\ref{fm:trafo}) of the response given the covariates.
\added{An advantage of marginalised multilevel models \citep{Heagerty_Zeger_2000}
over marginal transformation models is that the former models are
parameterised in terms of marginal effects of interest, whereas effect
shrinkage is part of the latter models. The distribution-free nature,
general applicability to all types of responses, and the relative computational
simplicity are, in our opinion, attractive features of transformation models compared to 
marginalised multilevel models.}

The models and estimation procedures introduced here are limited by some
practical and some conceptual constraints.  Response-varying regression
coefficients $\shiftparm(\ry)$ define distribution regression models
\citep{Foresi_Peracchi_1995,Chernozhukov_2013} where corresponding
mixed-effects models have been presented recently
\citep{Garcia_Marder_Wang_2018}.  This would be relatively straightforward
to implement in the framework presented here, in fact, stratification in
Weibull models was parameterised in a similar way.
A mix of continuous and
censored observations within one cluster would require to compute the
likelihood by partial integration over an $N_i$-dimensional normal, this is
currently not implemented.  On a more conceptual level, it seems impossible
to implement multilevel models for discrete or censored responses, because
the likelihood (\ref{fm:cll}) is only defined for contributions by
independent clusters.

\section*{Computational Details}

The empirical analyses presented in Sections~\ref{sec:applications} and
\ref{sec:sim} are
reproducible using the \code{mtram} package vignette \citep[Online Appendix,][]{vign:tram} in
package \pkg{tram} \citep{pkg:tram}.  Infrastructure for transformation
models from package \pkg{mlt} was used to define marginal models.  Augmented
Lagrangian Minimization implemented in the \code{auglag()} function of
package \pkg{alabama} \citep{pkg:alabama} was used for optimising the
log-likelihood.  Numerical integration to compute the discrete and censored
version of the log-likelihood was performed by \pkg{SparseGrid}
\citep{pkg:SparseGrid}.  GEEs were estimated using package \pkg{geepack}
\citep{pkg:geepack} and conditional mixed-effects (\ref{fm:normal} and
\ref{fm:tmME}) models using package
\pkg{tramME} \citep{pkg:tramME}. Packages \pkg{lme4} \citep{pkg:lme4} 
and \pkg{glmmTMB} \citep{pkg:glmmTMB} were used to fit generalised mixed-effects
models. All results were obtained using \textsf{R} version
4.2.2 \citep{R}.

\section*{Acknowledgements}

The authors would like to thank Leonhard Held, Thomas Kneib, Nadja Klein
and B{\'a}lint Tam{\'a}si for interesting discussions. 
Luisa Barbanti received a UZH Graduate Campus travel grant for a research stay in Berlin,
during which this paper was finalised.
Torsten Hothorn received funding from the Swiss National Science Foundation, 
grant number 200021\_184603.
 
\bibliography{mlt,packages}

\newpage

\begin{appendix}

\section{Likelihood Function: Censored and Discrete Case} \label{app:cens}

The $i$th contribution to the likelihood~(\ref{fm:dll})is given by the 
$N_i$-dimensional normal integral
\begin{eqnarray*}
\exp(\ell_i(\parm, \shiftparm, \varparm)) = 
   \int_{\zvec(\ubar{\yvec}_i \mid \parm, \shiftparm, \varparm)}^{\zvec(\bar{\yvec}_i \mid \parm, \shiftparm, \varparm)}
      \phivec_{N_i}\left(\zvec, \nullvec_{N_i}, \mU_i \mLambda(\varparm) \mLambda(\varparm)^\top \mU_i^\top + \mI_{N_i}\right) \, d\zvec.
\end{eqnarray*}
With $\mD_i(\varparm) = \text{diag}(\mU_i \mLambda(\varparm) \mLambda(\varparm)^\top \mU_i^\top + \mI_{N_i}) \cdot \mI_{N_i}$ 
we obtain the corresponding correlation matrix as 
\begin{eqnarray*}
\mC_i(\varparm) & = & \mD_i(\varparm)^{-\nicefrac{1}{2}} \mSigma_i(\varparm) \mD_i(\varparm)^{-\nicefrac{1}{2}} = \mV_i(\varparm) \mV_i(\varparm)^\top + \mD_i(\varparm)^{-1} \in \RR^{N_i \times N_i} \\
\mV_i(\varparm) & = & \mD_i(\varparm)^{-\nicefrac{1}{2}} \mU_i \mLambda(\varparm) \in \RR^{N_i \times \dimparmrand}
\end{eqnarray*}
and the integration limits become (again with $\zvec()$ defined in (\ref{fm:z}))
\begin{eqnarray*}
\ubar{\zvec} = \mD_i(\varparm)^{-\nicefrac{1}{2}} \zvec(\ubar{\yvec}_i \mid \parm, \shiftparm,
\varparm) \text{ and } 
\bar{\zvec} =
\mD_i(\varparm)^{-\nicefrac{1}{2}} \zvec(\bar{\yvec}_i \mid \parm, \shiftparm,
\varparm).
\end{eqnarray*}
According to \THcite{Marsaglia}{Marsaglia_1963}, the above normal probability can be written as
\begin{eqnarray*}
 \int_{\ubar{\zvec}}^{\bar{\zvec}} \phivec_{N_i}\left(\zvec, \nullvec_{N_i}, \mC_i(\varparm)\right) \, d\zvec
=   \int_{\RR^\dimparmrand} \phivec_\dimparmrand(\wvec, \nullvec_\dimparmrand, \mI_\dimparmrand) 
          \int_{\ubar{\zvec} - \mV \wvec}^{\bar{\zvec} - \mV \wvec} \phivec_{N_i}\left(\yvec, \nullvec_{N_i}, \mD_i(\varparm)^{-1}\right) \, d\wvec d\yvec
\end{eqnarray*}
and can, following \THcite{Genz and Bretz}{Genz_Bretz_2009} here, further be simplified to
\begin{eqnarray*}
  & = &   \int_{\RR^\dimparmrand} \phivec_\dimparmrand(\wvec, \nullvec_\dimparmrand, \mI_\dimparmrand) 
          \prod_{\imath = 1}^{N_i} \left[\Phi\left(\frac{\bar{z}_\imath - \sum_{r = 1}^\dimparmrand v_{\imath r} w_{r}}{\sqrt{d_i}} \right) - 
                                        \Phi\left(\frac{\ubar{z}_\imath - \sum_{r = 1}^\dimparmrand v_{\imath r} w_{r}}{\sqrt{d_i}} \right)\right] d\wvec \\
  & \stackrel{\wvec = \Phi^{-1}_\dimparmrand(\qvec)}{=} &   \int_{[0,1]^\dimparmrand} 
          \prod_{\imath = 1}^{N_i} \left[\Phi\left(\frac{\bar{z}_\imath - \sum_{r = 1}^\dimparmrand v_{\imath r} \Phi^{-1}(q_{r})}{\sqrt{d_\imath}} \right) - 
                                        \Phi\left(\frac{\ubar{z}_\imath - \sum_{r = 1}^\dimparmrand v_{\imath r} \Phi^{-1}(q_{r})}{\sqrt{d_\imath}} \right)\right] d\qvec.
\end{eqnarray*}
The elements $d_\imath$ are the diagonal elements of $\mD_i(\varparm)^{-1}$
and thus standardisation of $\rz$ and $\mU_i \mLambda(\varparm)$
cancel out in this case such that we get
\begin{eqnarray*}
  & = & 
  \int_{[0,1]^\dimparmrand} 
          \prod_{\imath = 1}^{N_i} \left[\Phi\left(\tilde{\bar{z}}_\imath - \sum_{r = 1}^\dimparmrand \tilde{v}_{\imath r} \Phi^{-1}(q_{r}) \right) - 
                                        \Phi\left(\tilde{\ubar{z}}_\imath - \sum_{r = 1}^\dimparmrand \tilde{v}_{\imath r} \Phi^{-1}(q_{r}) \right)\right] d\qvec
\end{eqnarray*}
with $\tilde{\bar{z}}_\imath$ and $\tilde{\ubar{z}}_\imath$ being the
elements of $\zvec(\bar{\yvec}_i \mid \parm, \shiftparm, \varparm)$ and
$\zvec(\ubar{\yvec}_i \mid \parm, \shiftparm, \varparm)$, respectively, and
$\tilde{v}_{\imath r}$ are the elements of $\mU_i \mLambda(\varparm)$.

The latter expression is an $\dimparmrand$-dimensional integral over the
unit cube (random intercept models have $\dimparmrand = 1$ and correlated
random intercept/random slope models correspond to $\dimparmrand = 3$) of
products of univariate normal probabilities.  It should be noted that,
unlike using an Laplace or other approximation of the likelihood, the above
term is the exact likelihood contribution.  It can be approximated up to any
desired accuracy using numerical integration procedures. An analytic
expression for the score function seems quite challenging and one thus has
to rely on numerical approaches such as sparse grids \citep{Heiss_Winschel_2008}.

\newpage
\section{Likelihood and Score Function: Continuous Case} \label{app:cont}

The joint probability of $\yvec_i \in \RR^{N_i}$ is given by:
\begin{eqnarray*}
\Prob(\mY_i \le \yvec_i \mid \mX_i, \mU_i) = \mPhi_{\nullvec_{N_i}, \mSigma_i(\varparm)}
\left( \mD_i(\varparm) \Phi^{-1}_{N_i} \left( \pZ_{N_i} \{ 
\mD_i(\varparm)^{-1} [\mA(\yvec_i)\parm - \mX_i \shiftparm] \}
\right) \right).
\end{eqnarray*}
To simplify the notation, we define $\zvec()$ as in (\ref{fm:z}):
\begin{eqnarray*} 
\zvec(\yvec_i \mid \parm, \shiftparm, \varparm) = \mD_i(\varparm) 
  \Phi^{-1}_{N_i}(\pZ_{N_i}\{\mD_i(\varparm)^{-1}[\mA(\yvec_i)\parm - \mX_i
\shiftparm]\}).
\end{eqnarray*} 
We can derive the corresponding joint density for an arbitrary $\pZ$:
\begin{eqnarray}
f_{\mY_i}(\yvec_i \mid \parm, \shiftparm, \varparm) 
& = & \left((2 \pi)^{N_i} \left| \mL_i(\varparm) \mL_i(\varparm)^\top
\right| \right)^{-\nicefrac{1}{2}} \times \nonumber \\
& & \quad \exp\left(-\frac{1}{2} \norm{\zvec(\yvec_i \mid \parm, \shiftparm,
\varparm)^\top \mL_i(\varparm)^{-1}}_2^2 \right) \times \nonumber \\
& & \quad \prod_{\imath=1}^{N_i} 
\frac{ \mD_i(\varparm)_{\imath \imath} 
       \dZ \{ (\mD_i(\varparm)^{-1})_{\imath \imath} [\avec(y_{i \imath})^\top \parm - \mX_i
\shiftparm] \}}{\dN(\Phi^{-1} \left( \pZ \{ 
\mD_i(\varparm)^{-1} [\avec(y_{i \imath})^\top \parm - \mX_i \shiftparm] \}
\right))} \,
(\mD_i(\varparm)^{-1})_{\imath \imath} \,
\avec^\prime(y_{i \imath})^\top \parm \nonumber \\
= & &  \left| \mL_i(\varparm) \mL_i(\varparm)^\top \right|^{-\nicefrac{1}{2}} \times \nonumber \\
& & \quad \exp\left(-\frac{1}{2} \norm{\zvec(\yvec_i \mid \parm, \shiftparm, \varparm)^\top \mL_i(\varparm)^{-1}}_2^2 \right) \times \nonumber \\
& & \quad  \exp\left(\frac{1}{2} \norm{\mD_i(\varparm)^{-1}
\zvec(\yvec_i \mid \parm, \shiftparm, \varparm)}_2^2\right)  \times \nonumber \\
& & \quad 
  \prod_{\imath=1}^{N_i} \dZ \{ (\mD_i(\varparm)^{-1})_{\imath \imath}
  [\avec(y_{i \imath})^\top \parm - \mX_i \shiftparm] \} \,
  \avec^\prime(y_{i \imath})^\top \parm \nonumber \\
= & &  \left| \mL_i(\varparm) \mL_i(\varparm)^\top \right|^{-\nicefrac{1}{2}} \times \nonumber \\
& & \quad 
  \exp\left(-\frac{1}{2} \zvec(\yvec_i \mid \parm, \shiftparm, \varparm)^\top 
  \mSigma_i(\varparm)^{-1} 
  \zvec(\yvec_i \mid \parm, \shiftparm, \varparm)
  \right) \times \nonumber \\
& & \quad 
  \exp\left(\frac{1}{2} \zvec(\yvec_i \mid \parm, \shiftparm, \varparm)^\top 
  \mD_i(\varparm)^{-2}
  \zvec(\yvec_i \mid \parm, \shiftparm, \varparm)
  \right) \times \nonumber \\
& & \quad 
  \prod_{\imath=1}^{N_i} \dZ \{ (\mD_i(\varparm)^{-1})_{\imath \imath}
  [\avec(y_{i \imath})^\top \parm - \mX_i \shiftparm] \} \,
  \avec^\prime(y_{i \imath})^\top \parm \nonumber.
\end{eqnarray}

The resulting log-likelihood contribution~(\ref{fm:cll}) for the $i$th observation is given by:
\begingroup
\allowdisplaybreaks
\begin{eqnarray*}
\ell_i(\parm, \shiftparm, \varparm) & \approx &
\log(f_{\mY_i}(\yvec_i \mid \parm, \shiftparm, \varparm)) \\
& = &
-\frac{1}{2}\log \left|\mL_i(\varparm)\mL_i(\varparm)^\top\right|
- \frac{1}{2} \norm{\zvec(\yvec_i \mid \parm, \shiftparm, \varparm)^\top \mL_i(\varparm)^{-1}}_2^2 + \\
& & \quad
\frac{1}{2} \norm{\mD_i(\varparm)^{-1} \zvec(\yvec_i \mid \parm, \shiftparm, \varparm)^\top}_2^2 + \\
& & \quad \log_{N_i}(\dZ_{N_i} \{ \mD_i(\varparm)^{-1} [\mA(\yvec_i) \parm - \mX_i \shiftparm] \} )^\top
\onevec_{N_i} + \\
& & \quad \log_{N_i}(
\mA^\prime(\yvec_i) \parm)^\top \onevec_{N_i} \\
& = & -\frac{1}{2}\log \left|\mSigma_i(\varparm)\right| + \\
& & - \frac{1}{2} \zvec(\yvec_i \mid \parm, \shiftparm, \varparm)^\top 
(\mSigma_i(\varparm)^{-1} - \mD_i(\varparm)^{-2})
\zvec(\yvec_i \mid \parm, \shiftparm, \varparm) + \\
& & \quad \log_{N_i}(\dZ_{N_i} \{ \mD_i(\varparm)^{-1} [\mA(\yvec_i) \parm - \mX_i \shiftparm] \} )^\top
\onevec_{N_i} + \\
& & \quad \log_{N_i}(
\mA^\prime(\yvec_i) \parm)^\top \onevec_{N_i}.
\end{eqnarray*}
\endgroup
The score function for all model parameters $\parm, \shiftparm$, and
$\varparm$ can be derived based on the results of \THcite{Stroup}{Stroup_2012} as applied to
normal linear mixed-effects models by \THcite{Wang and Merkle}{Wang_Merkle_2018}.

With the $\dimparmvar = \nicefrac{\dimparmrand (\dimparmrand + 1)}{2}$
unique elements $\varparm = (\gamma_1, \dots, \gamma_\dimparmvar)^\top$
of the lower Cholesky factor $\mLambda(\varparm)$ we get
\begin{eqnarray*}
\frac{\partial \mSigma_i(\varparm)}{\partial \gamma_m} =
  \mU_i \frac{\partial \mLambda(\varparm)\mLambda(\varparm)^\top}{\partial \gamma_m} \mU_i^\top
=
  \mU_i\left(\frac{\partial \mLambda(\varparm)}{\partial \gamma_m} \mLambda(\varparm)^\top +
             \mLambda(\varparm) \frac{\partial \mLambda(\varparm)^\top}{\partial \gamma_m}\right) \mU_i^\top.
\end{eqnarray*}
The derivative of $\mLambda(\varparm)$ with respect to an element $\gamma_m$
of $\varparm$ is a matrix of zeros with the exception of a single one at the position of
$\gamma_m$.
Moreover, we compute:
\begin{eqnarray*}
\frac{\partial \mD_i(\varparm)}{\partial \gamma_m} &=& 
\frac{1}{2} \mD_i(\varparm)^{-1} \text{diag} \left( 
\mU_i\left(\frac{\partial \mLambda(\varparm)}{\partial \gamma_m} \mLambda(\varparm)^\top + 
             \mLambda(\varparm) \frac{\partial \mLambda(\varparm)^\top}
             {\partial \gamma_m}\right) \mU_i^\top \right) \cdot \mI_{N_i} \\
&=& \frac{1}{2} \mD_i(\varparm)^{-1} \text{diag} \left( \frac{\partial \mSigma_i(\varparm)}
{\partial \gamma_m} \right) \cdot \mI_{N_i} \\
\frac{\partial \mD_i(\varparm)^{-1}}{\partial \gamma_m} &=& 
- \frac{1}{2} \left( \mD_i(\varparm)^{-1} \right)^3 
\text{diag} \left( \frac{\partial \mSigma_i(\varparm)}{\partial \gamma_m} \right) \cdot \mI_{N_i}  \\
\frac{\partial (\mD_i(\varparm)^{-1})^2}{\partial \gamma_m} &=& 
- \left( \mD_i(\varparm)^{-1} \right)^4 \text{diag} 
\left( \frac{\partial \mSigma_i(\varparm)}{\partial \gamma_m} \right) \cdot \mI_{N_i} \\ &=&
- \left( \text{diag}(\mSigma_i(\varparm)) \right)^{-2}
\text{diag} \left( \frac{\partial \mSigma_i(\varparm)}{\partial \gamma_m} \right) \cdot \mI_{N_i}
\end{eqnarray*}
and
\begingroup
\allowdisplaybreaks
\begin{eqnarray*}
\frac{\partial \zvec(\yvec_i \mid \parm, \shiftparm, \varparm)}{\partial \gamma_m} &=&
\frac{\partial \mD_i(\varparm) 
  \Phi^{-1}_{N_i}(\pZ_{N_i}\{\mD_i(\varparm)^{-1}[\mA(\yvec_i)\parm - \mX_i \shiftparm]\})}
  {\partial \gamma_m} \\
&=& \left( \frac{\partial}{\partial \gamma_m} \mD_i(\varparm) \right) 
\Phi^{-1}_{N_i}(\pZ_{N_i}\{\mD_i(\varparm)^{-1}[\mA(\yvec_i)\parm - \mX_i \shiftparm]\}) + \\
&\qquad +&  \mD_i(\varparm) \left( \frac{\partial}{\partial \gamma_m} 
\Phi^{-1}_{N_i}(\pZ_{N_i}\{\mD_i(\varparm)^{-1}[\mA(\yvec_i)\parm - \mX_i \shiftparm]\}) \right) \\
&=& \frac{1}{2} \mD_i(\varparm)^{-1} \text{diag} \left( \frac{\partial \mSigma_i(\varparm)}{\partial \gamma_m} \right) \cdot \mI_{N_i}
\Phi^{-1}_{N_i}(\pZ_{N_i}\{\mD_i(\varparm)^{-1}[\mA(\yvec_i)\parm - \mX_i \shiftparm]\}) + \\
&\qquad +&  \mD_i(\varparm) \frac{f_{N_i}\{\mD_i(\varparm)^{-1}[\mA(\yvec_i)\parm - \mX_i \shiftparm]\}}
{\dN_{N_i}[\Phi^{-1}_{N_i}(\pZ_{N_i}\{\mD_i(\varparm)^{-1}[\mA(\yvec_i)\parm - \mX_i \shiftparm]\})]}
[\mA(\yvec_i)\parm - \mX_i \shiftparm] \frac{\partial \mD_i(\varparm)^{-1}}{\partial \gamma_m}\\
&=& \frac{1}{2} \mD_i(\varparm)^{-2} \text{diag} \left( \frac{\partial \mSigma_i(\varparm)}{\partial \gamma_m} \right) \cdot \mI_{N_i} \zvec(\yvec_i \mid \parm, \shiftparm, \varparm) + \\
&\qquad -& \frac{1}{2}  \mD_i(\varparm)^{-2}  \text{diag} \left( \frac{\partial \mSigma_i(\varparm)}{\partial \gamma_m} \right) \cdot \mI_{N_i} \frac{f_{N_i}\{\mD_i(\varparm)^{-1}[\mA(\yvec_i)\parm - \mX_i \shiftparm]\}[\mA(\yvec_i)\parm - \mX_i \shiftparm] }
{\dN_{N_i}[\Phi^{-1}_{N_i}(\pZ_{N_i}\{\mD_i(\varparm)^{-1}[\mA(\yvec_i)\parm - \mX_i \shiftparm]\})]}
\\
&=& \frac{1}{2} \mD_i(\varparm)^{-2} \text{diag} \left( \frac{\partial \mSigma_i(\varparm)}{\partial \gamma_m} \right) \cdot \mI_{N_i} \times \\
&&\qquad \left[ \zvec(\yvec_i \mid \parm, \shiftparm, \varparm)
- \frac{f_{N_i}\{\mD_i(\varparm)^{-1}[\mA(\yvec_i)\parm - \mX_i \shiftparm]\}[\mA(\yvec_i)\parm - \mX_i \shiftparm] }
{\dN_{N_i}[\Phi^{-1}_{N_i}(\pZ_{N_i}\{\mD_i(\varparm)^{-1}[\mA(\yvec_i)\parm - \mX_i \shiftparm]\})]} \right]
\end{eqnarray*}
\endgroup

Thus
\begingroup
\allowdisplaybreaks
\begin{eqnarray*}
\frac{\partial \ell_i(\parm, \shiftparm, \varparm)}{\partial \gamma_m} & = & 
  -\frac{1}{2}\trace\left(\mSigma_i(\varparm)^{-1} \frac{\partial \mSigma_i(\varparm)}{\partial \gamma_m}
\right) + \\
& & -\frac{1}{2} \Bigg[ \left(
\frac{\partial \zvec(\yvec_i \mid \parm, \shiftparm, \varparm)^\top}{\partial \gamma_m} \right)
(\mSigma_i(\varparm)^{-1} - \mD_i(\varparm)^{-2}) 
\zvec(\yvec_i \mid \parm, \shiftparm, \varparm) + \\ & & \qquad \qquad
\zvec(\yvec_i \mid \parm, \shiftparm, \varparm)^\top
\left(\frac{\partial }{\partial \gamma_m} (\mSigma_i(\varparm)^{-1} - 
\mD_i(\varparm)^{-2}) \right)
\zvec(\yvec_i \mid \parm, \shiftparm, \varparm) + \\ & & \qquad \qquad
\zvec(\yvec_i \mid \parm, \shiftparm, \varparm)^\top
(\mSigma_i(\varparm)^{-1} - \mD_i(\varparm)^{-2})
\left(\frac{\partial }{\partial \gamma_m} \zvec(\yvec_i \mid \parm, \shiftparm, \varparm) \right)
\Bigg] + \\ & & \, +
\frac{\dZ_{N_i}^\prime(\mD_i(\varparm)^{-1}[\mA(\yvec_i) \parm - \mX_i \shiftparm])}
{\dZ_{N_i}(\mD_i(\varparm)^{-1}[\mA(\yvec_i) \parm - \mX_i \shiftparm])}
[\mA(\yvec_i) \parm - \mX_i \shiftparm] 
\frac{\partial \mD_i(\varparm)^{-1}}{\partial \gamma_m} \\
& = & 
  -\frac{1}{2}\trace\left(\mSigma_i(\varparm)^{-1} \frac{\partial \mSigma_i(\varparm)}{\partial \gamma_m}
\right) + \\
& & -\frac{1}{2} \Bigg[ \left(
\frac{\partial \zvec(\yvec_i \mid \parm, \shiftparm, \varparm)^\top}{\partial \gamma_m} \right)
(\mSigma_i(\varparm)^{-1} - \mD_i(\varparm)^{-2}) 
\zvec(\yvec_i \mid \parm, \shiftparm, \varparm) + \\ & & \qquad \qquad
- \zvec(\yvec_i \mid \parm, \shiftparm, \varparm)^\top
\Big(\mSigma_i(\varparm)^{-1} \frac{\partial \mSigma_i(\varparm)}{\partial \gamma_m}
\mSigma_i(\varparm)^{-1} + \\
& & \hspace{3cm}
\left( \mD_i(\varparm)^{-1} \right)^{4}
\text{diag} \left( \frac{\partial \mSigma_i(\varparm)}{\partial \gamma_m} \right) \cdot
\mI_{N_i} \Big)
\zvec(\yvec_i \mid \parm, \shiftparm, \varparm) + \\ & & \qquad \qquad
\zvec(\yvec_i \mid \parm, \shiftparm, \varparm)^\top
(\mSigma_i(\varparm)^{-1} - \mD_i(\varparm)^{-2})
\left(\frac{\partial \zvec(\yvec_i \mid \parm, \shiftparm, \varparm)}{\partial \gamma_m}  \right)
\Bigg] + \\ & & \quad +
\frac{\dZ_{N_i}^\prime(\mD_i(\varparm)^{-1}[\mA(\yvec_i) \parm - \mX_i \shiftparm])}
{\dZ_{N_i}(\mD_i(\varparm)^{-1}[\mA(\yvec_i) \parm - \mX_i \shiftparm])}
[\mA(\yvec_i) \parm - \mX_i \shiftparm] 
\frac{\partial \mD_i(\varparm)^{-1}}{\partial \gamma_m} \\
\frac{\partial \ell_i(\parm, \shiftparm, \varparm)}{\partial \shiftparm} 
& = &
\zvec(\yvec_i \mid \parm, \shiftparm, \varparm)^\top (\mSigma_i(\varparm)^{-1} - 
\mD_i(\varparm)^{-2} ) \times \\ & & \quad
\frac{\dZ_{N_i}(\mD_i(\varparm)^{-1}[\mA(\yvec_i) \parm - \mX_i \shiftparm])}
{\dN_{N_i}(\Phi^{-1}_{N_i}(\pZ_{N_i}\{\mD_i(\varparm)^{-1}[\mA(\yvec_i)\parm - \mX_i \shiftparm]\}))}
\mX_i + \\
& & \quad
- \onevec_{N_i}^\top \left( 
\frac{f^\prime_{N_i}(\mD_i(\varparm)^{-1}[\mA(\yvec_i)\parm - \mX_i \shiftparm])}
{\dZ_{N_i}(\mD_i(\varparm)^{-1}[\mA(\yvec_i)\parm - \mX_i \shiftparm])} \mD_i(\varparm)^{-1} \mX_i \right) \\
\frac{\partial \ell_i(\parm, \shiftparm, \varparm)}{\partial \parm} 
& = &
- \zvec(\yvec_i \mid \parm, \shiftparm, \varparm)^\top (\mSigma_i(\varparm)^{-1} - 
\mD_i(\varparm)^{-2} ) \times \\ & & \quad
\frac{\dZ_{N_i}(\mD_i(\varparm)^{-1}[\mA(\yvec_i) \parm - \mX_i \shiftparm])}
{\dN_{N_i}(\Phi^{-1}_{N_i}(\pZ_{N_i}\{\mD_i(\varparm)^{-1}[\mA(\yvec_i)\parm - \mX_i \shiftparm]\}))}
\mA(\yvec_i) + \\ & & \quad
\onevec_{N_i}^\top \left( 
\frac{f^\prime_{N_i}(\mD_i(\varparm)^{-1}[\mA(\yvec_i)\parm - \mX_i \shiftparm])}
{\dZ_{N_i}(\mD_i(\varparm)^{-1}[\mA(\yvec_i)\parm - \mX_i \shiftparm])} \mD_i(\varparm)^{-1} \mA(\yvec_i) \right) +
\onevec_{N_i}^\top \frac{1}{\mA^\prime(\yvec_i) \parm} \mA^\prime(\yvec_i).
\end{eqnarray*}
\endgroup
For $\pZ = \Phi$, $\zvec(\yvec_i \mid \parm, \shiftparm, \varparm) = \mA(\yvec_i)\parm - \mX_i
\shiftparm$, the joint distribution simplifies to
\begin{eqnarray*}
\Prob(\mY_i \le \yvec_i \mid \mX_i, \mU_i) = \mPhi_{\nullvec_{N_i}, \mSigma_i(\varparm)}
\left( \mA(\yvec_i)\parm - \mX_i \shiftparm \right)
\end{eqnarray*}
and the joint density becomes:
\begin{eqnarray*}
f_{\mY_i}(\yvec_i \mid \parm, \shiftparm, \varparm)
& = & \left((2 \pi)^{N_i} \left| \mL_i(\varparm) \mL_i(\varparm)^\top
\right| \right)^{-\nicefrac{1}{2}} \times \nonumber \\
& & \quad 
  \exp\left(-\frac{1}{2} (\mA(\yvec_i)\parm - \mX_i \shiftparm)^\top 
  \mSigma_i(\varparm)^{-1} (\mA(\yvec_i)\parm - \mX_i \shiftparm)
  \right) \times \nonumber \\
& & \quad \prod_{\imath=1}^{N_i} \avec^\prime(y_{i \imath})^\top \parm.
\end{eqnarray*}
We obtain the corresponding log-likelihood:
\begin{eqnarray*}
\ell_i(\parm, \shiftparm, \varparm) & \approx & 
\log(f_{\mY_i}(\yvec_i \mid \parm, \shiftparm, \varparm)) \\
& \propto &
-\frac{1}{2}\log \left|\mSigma_i(\varparm)\right| + \\
& & \quad  - \frac{1}{2} (\mA(\yvec_i)\parm - \mX_i \shiftparm)^\top 
\mSigma_i(\varparm)^{-1} (\mA(\yvec_i)\parm - \mX_i \shiftparm) + \\
& & \quad \log_{N_i}(
\mA^\prime(\yvec_i) \parm)^\top \onevec_{N_i} \\
& = &
-\frac{1}{2}\log \left|\mSigma_i(\varparm)\right| 
- \frac{1}{2} \parm^\top\mA(\yvec_i)^\top \mSigma_i(\varparm)^{-1}
\mA(\yvec_i)\parm + \parm^\top\mA(\yvec_i)^\top \mSigma_i(\varparm)^{-1}
\mX_i \shiftparm + \\
& & - \frac{1}{2} \shiftparm^\top \mX_i^\top \mSigma_i(\varparm)^{-1} \mX_i \shiftparm
+   \log_{N_i}(\mA^\prime(\yvec_i) \parm)^\top \onevec_{N_i}.
\end{eqnarray*}
The scores are:
\begin{eqnarray*}
\frac{\partial \ell_i(\parm, \shiftparm, \varparm)}{\partial \gamma_m} & = & 
  -\frac{1}{2}\trace\left(\mSigma_i(\varparm)^{-1} \frac{\partial \mSigma_i(\varparm)}{\partial \gamma_m}
\right) + \\ 
& & \frac{1}{2} (\mA(\yvec_i)\parm - \mX_i \shiftparm)^\top \mSigma_i(\varparm)^{-1}
\frac{\partial \mSigma_i(\varparm)}{\partial \gamma_m} \mSigma_i(\varparm)^{-1} (\mA(\yvec_i)\parm - \mX_i \shiftparm) \\
\frac{\partial \ell_i(\parm, \shiftparm, \varparm)}{\partial \shiftparm} & = & 
\parm^\top\mA(\yvec_i)^\top \mSigma_i(\varparm)^{-1}
\mX_i - \shiftparm^\top \mX_i^\top \mSigma_i(\varparm)^{-1} \mX_i \\
\frac{\partial \ell_i(\parm, \shiftparm, \varparm)}{\partial \parm} & = & 
- \parm^\top\mA(\yvec_i)^\top \mSigma_i(\varparm)^{-1}
\mA(\yvec_i) + \shiftparm^\top \mX_i^\top \mSigma_i(\varparm)^{-1} \mA(\yvec_i)
+ \onevec_{N_i}^\top \frac{1}{\mA^\prime(\yvec_i) \parm} \mA^\prime(\yvec_i).
\end{eqnarray*}

\end{appendix}

\end{document}